\documentclass[twocolumn]{aa}  

\usepackage{graphicx}
\usepackage{subcaption}
\usepackage[allcolors=blue, hidelinks]{hyperref}
\usepackage{gensymb}
\usepackage[separate-uncertainty=true]{siunitx}
\usepackage{txfonts}
\begin{document}

   \title{General relativistic effects and the near-infrared and X-ray variability of Sgr A* I}

   \author{     Sebastiano D. von Fellenberg\inst{1}
          \and Gunther Witzel\inst{1}
          \and Michi Baub{\"o}ck\inst{2}
          \and Hui-Hsuan Chung\inst{1}
          \and Nicol{\'a}s Aimar\inst{3}
          \and Matteo Bordoni\inst{4}
          \and Antonia Drescher\inst{4}
          \and Frank Eisenhauer\inst{4}
          \and Reinhard Genzel\inst{4}
          \and Stefan Gillessen\inst{4}
          \and Nicola Marchili\inst{6,1}
          \and Thibaut Paumard\inst{3}
          \and Guy Perrin\inst{3}
          \and Thomas Ott \inst{4}
          \and Diogo Ribeiro \inst{4}
          \and Eduardo Ros \inst{1}
          \and Fr{\'e}d{\'e}ric Vincent\inst{3}
          \and Felix Widmann \inst{4}
          \and S. P.\ Willner \inst{5} 
          \and J. Anton Zensus \inst{1}
          }
   \institute{Max Planck Institute for Radio Astronomy, Auf dem H{\"u}gel 69, 53121 Bonn, Germany,
    \and University of Illinois Urbana-Champaign, 1002 W. Green Street, Urbana, IL 61801, USA  
    \and LESIA, Observatoire de Paris, Universit\'e PSL, CNRS, Sorbonne Universit\'e, Universit\'e de Paris, 5 place Jules Janssen, 92195 Meudon, France
    \and Max Planck Institute for Extraterrestrial Physics, Gie{\ss}enbachstra{\ss}e 1, 85748 Garching bei M{\"u}nchen, Germany
    \and Center for Astrophysics \textbar\ Harvard \& Smithsonian, 60 Garden St., Cambridge, MA 02138, USA
    \and Italian ALMA Regional Centre, INAF-Istituto di Radioastronomia, Via P. Gobetti 101, 40129 Bologna, Italy}

   \date{Received \today; accepted \today}

 
  \abstract{The near-infrared (NIR) and X-ray emission of Sagittarius A* shows occasional bright flares that are assumed to originate from the innermost region of the accretion flow. We identified $25$ $\SI{4.5}{\micro \meter}$ and $24$ X-ray flares in archival data obtained with the \textit{Spitzer} and \textit{Chandra} observatories. With the help of general relativistic ray-tracing code, we modeled trajectories of ``hot spots'' and studied the light curves of the flares for signs of the effects of general relativity. Despite their apparent diversity in shape, all flares share a common, exponential impulse response, a characteristic shape that is the building block of the variability. This shape is symmetric, that is, the rise and fall times are the same. Furthermore, the impulse responses in the NIR and X-ray are identical within uncertainties, with an exponential time constant $\tau\sim \SI{15}{\minute}$. The observed characteristic flare shape is inconsistent with hot-spot orbits viewed edge-on. Individually modeling the light curves of the flares, we derived constraints on the inclination of the orbital plane of the hot spots with respect to the observer ($i \sim \SI{30}{\degree}, < \SI{75}{\degree}$) and on the characteristic timescale of the intrinsic variability (tens of minutes). }

   \maketitle
%

\section{Introduction}
The Galactic Center massive black hole Sagittarius A* \citep[Sgr~A*;][]{Genzel2010,Morris2012} is one of the most studied astrophysical objects. Despite that, the mechanisms behind its emission are remarkably poorly understood \citep{Dodds-Eden2010}. While the steady emission in the radio is attributed to an outflow \citep[e.g.,][]{Brinkerink2016}, and the submillimeter emission is thought to originate from an accretion flow well described by semi-analytical       radiationally inefficient accretion flow models or jet models \citep[e.g.,][]{Yuan2003, Falcke2000}, the erratic flaring activity in the near infrared (NIR) remains a puzzle \citep{GravityCollaboration2021_xrayflare, 2018ApJ...863...15W, Witzel2021, Ponti2017, Eckart2012}. 

The substantial effort in modeling the submillimeter and radio emission with simulations, so-called general relativistic magnetohydrodynamic (GRMHD) simulations, that build on earlier semi-analytical work \citep{Dexter2010, Moscibrodzka2013, Chan2015, Davelaar2018, Dexter2020_model} has led to considerable success. 
The observed radio to submillimeter spectral energy distribution \cite[SED; e.g.,][]{Brinkerink2015, Bower2018, Liu2016, VonFellenberg2018, Bower2019} is well matched by these simulations, as are the variability \citep[e.g.,][]{Dexter2014} and the polarization \citep[e.g.,][]{Bower2015}.
Recently, the Event Horizon Telescope (EHT) Collaboration achieved the first image of the black hole shadow \citep{Falcke2000_bhshadow, eht_sgra_I, eht_sgra_II, eht_paper_III, eht_paper_IV, eht_paper_V, eht_paper_VI}. While the predicted morphology of the black hole shadow and innermost accretion flow seem consistent with data \citep[e.g.,][]{Lu2018}, it is unclear if the turbulent accretion flow scenario -- the standard and normal evolution (SANE) -- and the magnetically arrested accretion flow scenario -- the magnetically arrested disk (MAD) --  suffice to describe the accretion flow. For instance, \cite{Ressler2020} demonstrated that wind-fueled accretion leads to a low angular momentum flow with an inner MAD region.

Despite these successes, the fast NIR and X-ray variability is still not understood. It is generally accepted that the emission must originate from some form of nonthermal process \citep{YusefZadeh2006, Dodds-Eden2010, GRAVITYCollaboration2020flux, Witzel2021}. 
The flux density, spectral slopes, and temporal evolution of simultaneously and individually observed NIR and X-ray flares suggest that they are causally connected and originate from a localized region of the accretion flow \citep{Genzel2010, Barriere2014, Ponti2017, GravityCollaboration2021_xrayflare}. If an X-ray flares occurs, it is always accompanied by a NIR flare, but the reverse is not true \citep[e.g.,][]{Dodds-Eden2009}. A detailed analysis by \cite{Boyce2019, Boyce2021} of all available multiwavelength NIR and X-ray observations could not establish a significant temporal lead or delay of simultaneous NIR and X-ray flares. The extension of NIR and X-ray flares to the (sub)millimeter and radio regimes is difficult. While numerous studies have found tentative evidence in favor of a temporally delayed extension to longer wavelengths \citep[e.g.,][]{YusefZadeh2006, Yusef-Zadeh2009, Eckart2009, Witzel2021, Boyce2022, Wardle2021}, the correlation in total intensity measurements remains difficult to establish. Recent X-ray and polarimetric millimeter observations by \cite{Wielgus2022}, however, seem to confirm a delay of $\sim 30$ minute  between a bright X-ray flare and an increase in millimeter flux.
Several candidate mechanisms have been proposed: relativistic lensing and boosting \citep{Genzel2003, Broderick2006_hotspot, Hamaus2009, Karsen2017}, turbulent heating \citep{Comisso2020, Werner2021, Nattila2021}, magnetic reconnection \citep{Yuan2003, Yuan2009, Dodds-Eden2010, Mao2017, Chatterjee2020, Ripperda2020, Dexter2020_flare, Porth2021, Ripperda2022}, gap discharges \citep{Chen2018, Crinquand2020}, shocks \citep{Dexter2014}, and even more exotic scenarios such as tidal disruptions of asteroids \citep{Zubovas_2012}. GRMHD simulations tailored to study the accretion flow cannot generate emission from nonthermal (accelerated) electrons, as magnetohydrodynamics assumes a thermalized and collisional plasma. To include nonthermal emission, collisionless plasma is required, which is currently being studied in first-principle general relativistic (GR) particle in cell simulations \citep{Bransgrove2021,Galishnikova2022, Crinquand2022}.
The magnetic reconnection scenario has gained traction as a plausible emission mechanism, thought to occur frequently in MAD accretion flows \citep{Dexter2020_model, Porth2021}. Simulations by \cite{Ripperda2020, Ripperda2022} showed that reconnection can create and fill large vertical flux tubes with accelerated electrons. They are confined to the vertical magnetic field and thus orbit in the accretion disk. 
Relativistic effects (i.e., lensing and boosting) and electron acceleration mechanisms (i.e., turbulent heating, magnetic reconnection, shocks, and tidal disruption) both lead to variability in the observed emission. They, however, belong to two different categories: relativistic effects affect the direction and geodesic path of the photons regardless of the emission mechanism. Given that the NIR and X-ray emission of Sgr~A* likely originates from the direct vicinity of the black hole, the contribution to the observed emission may be significant.

The importance of dynamical timescales and GR effects in the context of Sgr~A* has long been disputed. The first NIR light curve of Sgr~A*, reported by \cite{Genzel2003}, showed substructure on a timescale of $\SI{20}{\minute}$ close to the orbital period of the innermost stable circular orbit (ISCO) of a $4 \cdot 10^6 \;\textup{M}_\odot $ black hole. This triggered the idea of identifying the substructure in the light curves with a ``hot spot'' in the accretion flow and using this orbital ``clock'' as a probe to test the black hole's gravitational potential \citep{Broderick2006_hotspot, Genzel2010}. However, the light curves alone did not show evidence for periodicity at any timescale \citep{Do2009}.

With the advent of the very large telescope interferometer (VLTI) GRAVITY, mapping the position of Sgr~A* and its progression during a flare became possible. In 2018, GRAVITY observed clockwise motion during three bright flares \citep{GRAVITYCollaboration2018_orbital, GravityCollaboration2020_orbital}. Furthermore, by tracking the linear polarization of Sgr~A*'s emission, the GRAVITY measurements could demonstrate a characteristic polarization loop in the Q-U plane, as expected for a hot spot moving close to the ISCO \citep{GravityCollaboration2020_polariflares}. Recently, \cite{Wielgus2022} suggested a similar Q-U loop to be present in submillimeter light curves using ALMA observations. Thus, while different aspects and implications of the hot-spot picture are far from established or understood, there is now considerable evidence in favor of it.

The results from the GRAVITY flare observations, the EHT image, and the ALMA polarization measurement have placed constraints on the geometry of the system. Based on the flare motion, light curve contrast, and polarization loops, \cite{GRAVITYCollaboration2018_orbital, GravityCollaboration2020_polariflares, GravityCollaboration2020_orbital} favored low inclination, in line with the modeling of the ALMA polarization measurements of \cite{Wielgus2022}. Similarly, the comparison of the radio image of Sgr A* with GRMHD models suggested a low inclination.

In this Letter we aim to identify and constrain relativistic effects in the light curve of Sgr~A* in the NIR and X-ray observing bands. In particular, we try to disentangle the relativistic modulation of the light curves from the intrinsic emission of the nonthermal electrons. While we also show the results of a periodicity analysis, we -- in contrast to past studies, such as \cite{Do2009} -- do not focus on finding evidence of periodicity in the data to confirm or rule out the orbiting hot-spot model. Instead, we assume the hot-spot model to be a valid description of the NIR and X-ray emission of the accretion flow. Using the GR model developed in the context of \cite{GravityCollaboration2020_orbital}, we try to establish constraints on the intrinsic variability as well as the fundamental system parameters, such as the inclination and orbital radius. 

\section{Data}
In order to obtain a representative set of NIR and X-ray flares of Sgr~A*, we used the well-characterized data sets obtained by \textit{Spitzer}/IRAC and the \textit{Chandra} X-ray Observatory from \cite{Hora2014}, \cite{Witzel2018}, and \cite{Witzel2021}. Details on the \textit{Chandra} data reduction are available in \cite{2019ApJ...875...44Z}.

The \textit{Spitzer}/IRAC observations were conducted in the M band ($\lambda_{\rm{cen.}}=\SI{4.5}{\micro \metre}$) and consist of eight light curves with $\sim24$ hours of continuous observations each. These light curves provide data with homogeneous sampling ($\SI{0.6}{\second}$ cadence) and measurement uncertainties, and they have signal-to-noise properties for detecting variability similar to data from the Keck/NIRC2 or VLT/NACO instruments at $\SI{2.2}{\micro \metre}$. However, they have been obtained via differential photometry and do not provide absolute flux density measurements with the accuracy of the ground-based telescopes. Nevertheless, for bright flaring states they offer a reliable determination of Sgr~A*'s variability in the NIR. Here, we re-binned the data to one-minute cadence. The typical uncertainty in the observed light curve is on the order of $\SI{\pm0.2}{mJy}$ or $\SI{\pm0.5}{mJy}$ for the extinction-corrected data \citep[assuming $A_M = 1.00 \pm 0.14$~mag;][]{Fritz2011}. 

\begin{figure*}[t]
    \centering
    \includegraphics[width=0.985\textwidth]{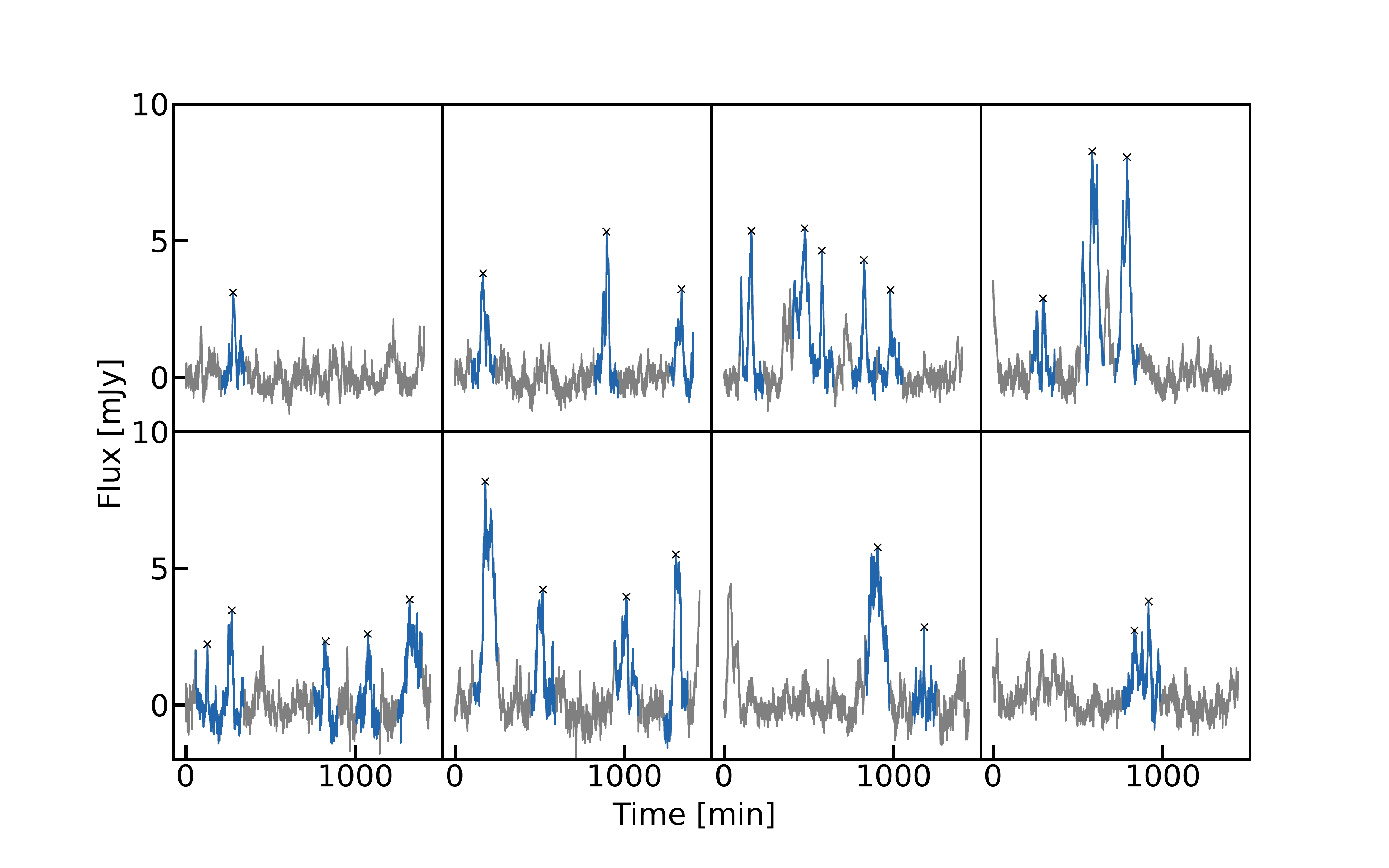}
    \caption{All eight \textit{Spitzer} epochs plotted alongside one another. The light curves \citep{Witzel2021} have a typical duration of $24$ hours and have been binned to one minute. The blue highlighted parts of the light curve show the $25$ segments of the light curve in which the observed (i.e., not de-reddened) flux density rose above $\SI{2}{\milli Jy}$, and the cross marks show the peak of a flare segment.}
    \label{fig:flare_overview}
\end{figure*}

The concept of a ``flare'' is heuristic, and so far no efforts have established a clear definition of what constitutes a ``flaring'' state. Based on the flux distribution, which starts to deviate from a log-normal distribution at a flux density of $\SI{\sim3}{mJy}$ in the K band \citep{GRAVITYCollaboration2020flux}, we defined all variability excursions brighter than $F_{\nu, \rm{obs}} \geq \SI{2}{mJy} = F_{\nu, \rm{de-redden}} \SI{5}{mJy}$ as flares. It will become clear in the following analysis that such a definition, even if purely phenomenological, can provide the basis for a time-domain characterization of the variability. For these flux excursions, we defined the highest flux point as the midpoint and selected a window of $\pm \SI{70}{\minute}$ around it. This ensured that, in most cases, the entire bright state of a flare was entirely covered when individual flaring episodes were separated out. Additionally, the interval of 140 minutes is much longer than the anticipated ISCO timescale around the black hole, which ensured that we did not miss any interesting features during the flare. \autoref{fig:flare_overview} gives an overview of all eight \textit{Spitzer}/IRAC observation epochs, during which $25$ such flares occurred.

We proceeded in a similar way to identify X-ray flares in the \textit{Chandra} data. We used all available data from the ACIS-I, ACIS-S, and ACIS-S/HETG arrays through 2017, including data that showed low-level contamination from the magnetar outburst \citep{Eatough2013}. We selected flares in a similar fashion as in the \textit{Spitzer} data set, selecting windows of \SI{\pm70}{\minute} around peaks in the light curve with significant flux. We found $26$ segments of the light curves with X-ray flux densities above the threshold. 

\section{Analysis and results}
\subsection{The characteristic response function of the Sgr~A* variability}\label{sec:pca_shape}
In order to determine the time-domain characteristics of NIR and X-ray flares, and to inform their physically motivated modeling, we normalized all flares in our sample and shifted them such that their peak was centered at $t=\SI{0}{\minute}$. Stacking and normalizing reveals a characteristic profile (\autoref{fig:arch_flare}) for both the NIR and X-ray flares. We refer to it as the impulse response function.

\subsubsection{\textit{Spitzer} data}
\begin{figure*}[t]
    \centering
    \includegraphics[width=0.985\textwidth]{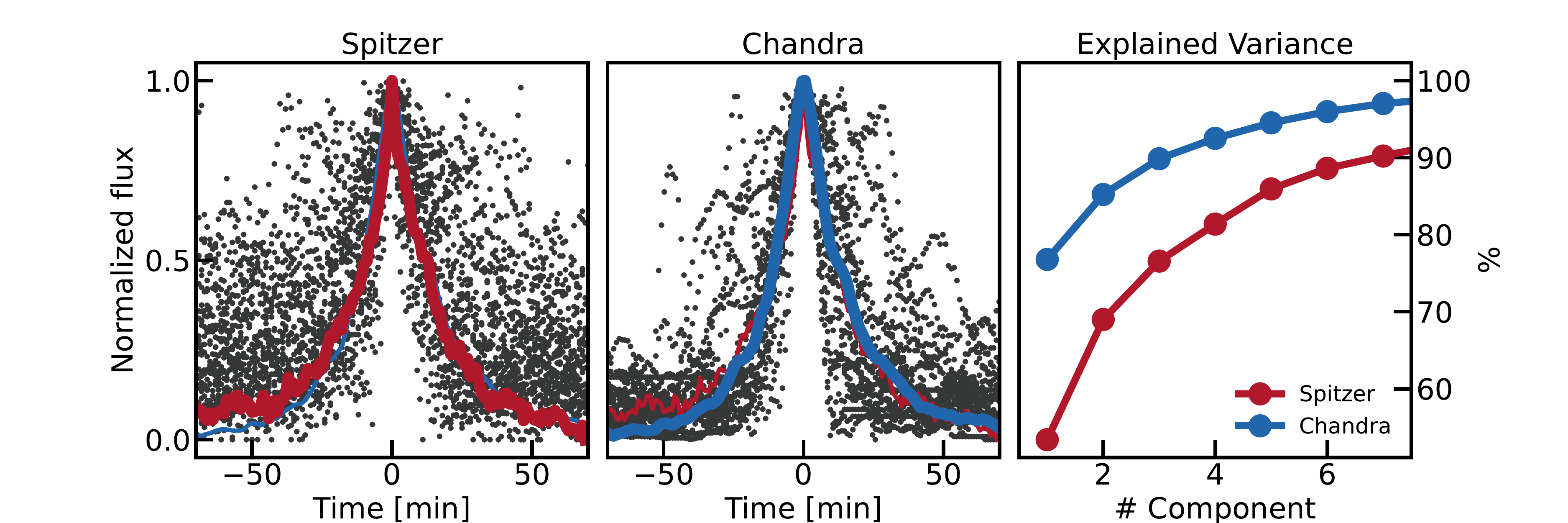}
    \caption{Characteristic flare shape of near infrared and X-ray flares. Left: Normalized data segments for the $25$ flares highlighted in \autoref{fig:flare_overview}. The thick red line indicates the first principal component derived from the data and the thin blue line the first principal component derived from the X-ray \textit{Chandra} data (see the middle panel).
    Middle: Same as the left panel for the $26$ \textit{Chandra} X-ray flares. Right: Explained cumulative variance of the PCA components of \textit{Spitzer} (red) and \textit{Chandra} data (blue).}
    \label{fig:arch_flare}
\end{figure*}

\begin{figure}
    \centering
    \includegraphics[width=0.485\textwidth]{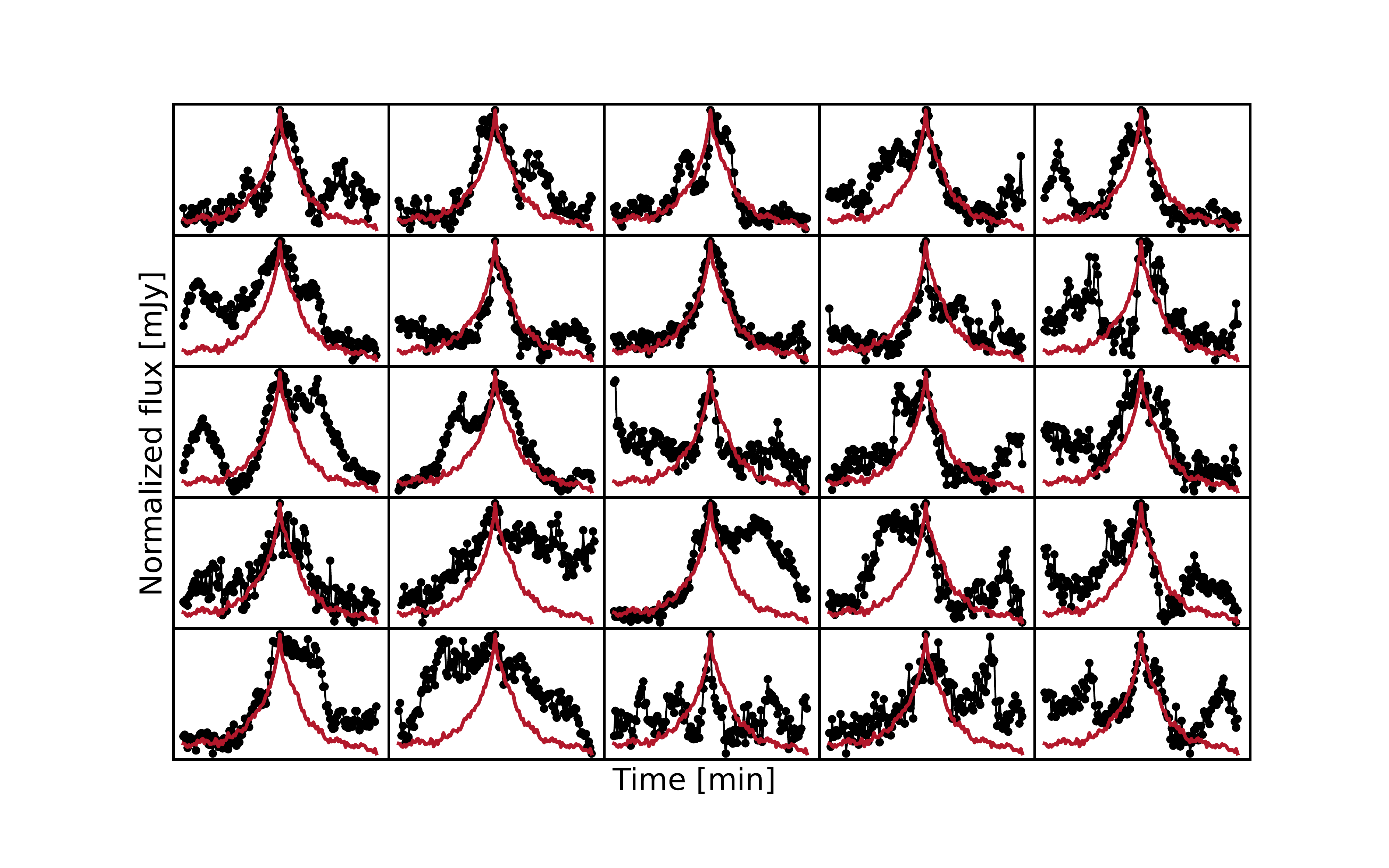}
    \caption{All $25$ normalized \textit{Spitzer} flares (black points), together with the first component of the PCA (thin red line).}
    \label{fig:pca_flares}
\end{figure}

To extract the characteristic response function, we decomposed the data using a principal component analysis \citep[PCA;][]{Pearson1901}. \autoref{fig:arch_flare} shows the first PCA component of the NIR and X-ray data. In the NIR data it can explain $53\%$ of the observed variance and matches the described average flare shape. \autoref{fig:pca_flares} illustrates the first component for all $25$ \textit{Spitzer} flares\footnote{For illustration purposes, both the component and the flares have been normalized; normalization is, however, not necessary for the derivation of the components.}. The overall shape of each flare is well described by the first PCA component, but only a few flares are fully described by it. Many flares have secondary side peaks, and some are much broader. Nevertheless, the derived profile seems to serve as an elementary building block of the variability.

\subsubsection{\textit{Chandra} data}
A very similar characteristic shape is found for the X-ray data, and it matches the NIR shape almost perfectly (compare the thin red line to thick blue line in \autoref{fig:arch_flare}). Overall, the X-ray flares seem to be less complex, consistent with $\sim75\%$ of the variance explained in the first component. 

\subsection{An exponential response function}
\begin{figure}
    \centering
    \includegraphics[width=0.485\textwidth]{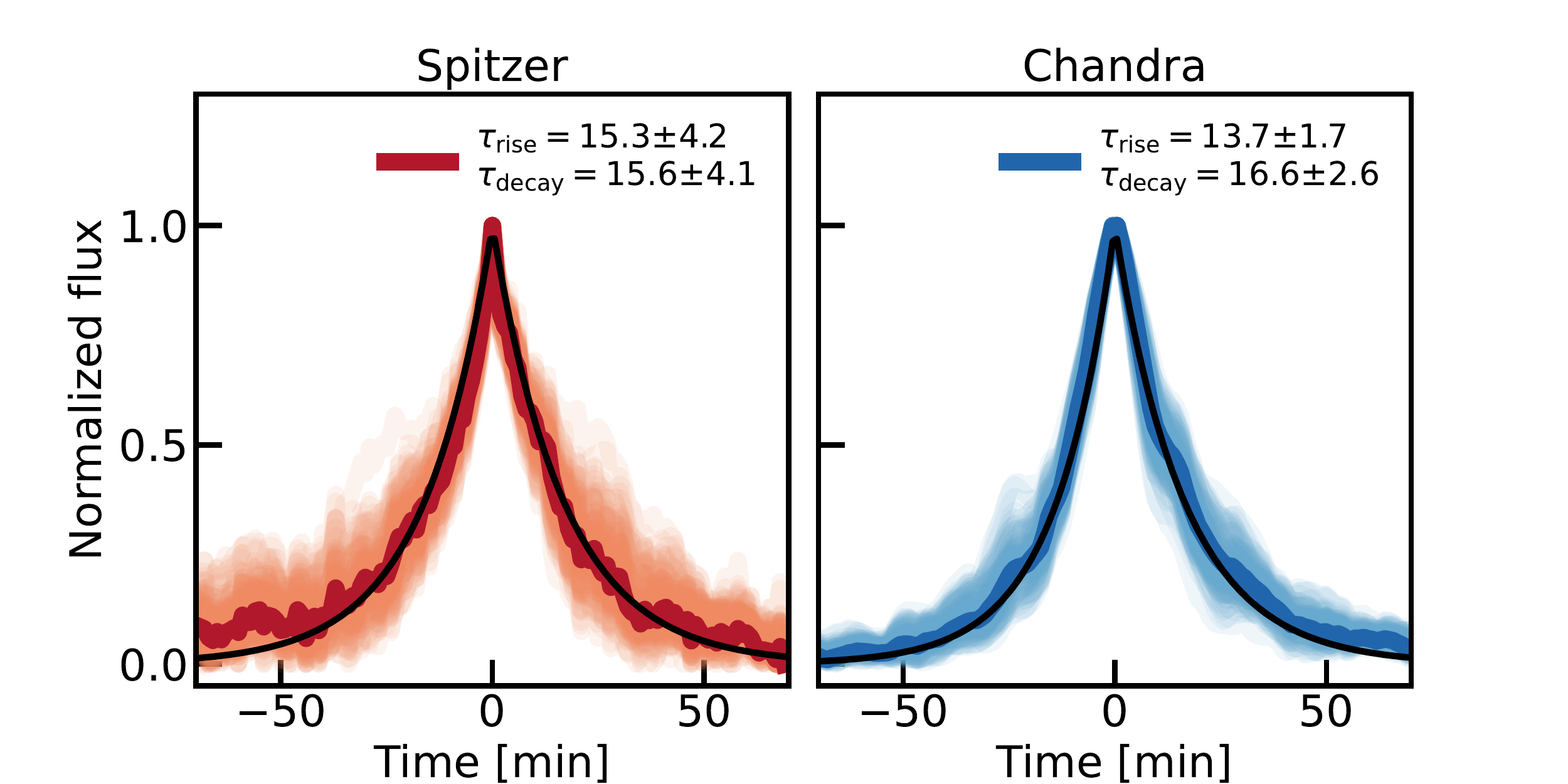}
    \caption{Exponential rise and decay time of the characteristic flare shape. Left: Principal component derived from the \textit{Spitzer} data. The light red lines show the first components derived from $100$ bootstrapped flare data sets. The black line shows the best fit exponential rise and decay. The uncertainty denoted in the figure legend is derived from the standard deviation of the $\tau$ value in the bootstrap samples. Right: Same but for the \textit{Chandra} X-ray data.}
    \label{fig:rise_decay_time}
\end{figure}
The derived response function is well described by an exponential rise and decay. In order to determine the rise and decay times, we fitted the NIR and X-ray PCA component\footnote{We allowed for a constant offset of the exponential function in order to avoid biasing the rise and decay time by the noise level in the first PCA component; see \autoref{sec:pca_noise_bias} for details.} with an exponential with a free $\tau_{\rm{rise/decay}}$. We found a NIR $\tau_{\rm{rise}} = \SI{15.3\pm4.2}{\minute}$ and $\tau_{\rm{decay}} = \SI{15.6\pm4.1}{\minute}$ (left \autoref{fig:rise_decay_time}). 
For the X-ray, we found $\tau_{\rm{rise}} = \SI{13.7\pm1.7}{\minute}$ and $\tau_{\rm{decay}} = \SI{16.6\pm2.6}{\minute}$ (right \autoref{fig:rise_decay_time}).
The uncertainty was derived by calculating the standard deviation of $100$ bootstrapped \citep[as described in][]{PresTeukVettFlan92} surrogate PCAs, as indicated in \autoref{fig:rise_decay_time}. 
There is no indication of different rise and fall times in the NIR. The rise time in the X-ray is about $\SI{3}{\minute}$ shorter than the decay time, but this difference is not significant. Further, the rise and decay times in the two bands are consistent with each other.

In summary, a simple, symmetric exponential rise and decay with a characteristic time of $\tau\approx\SI{15}{\minute}$ describes $50\%$ and $75\%$ of the variance in the NIR and X-ray light curves, respectively, for the time $\SI{\pm70}{\minute}$ before and after the flare peak. 

\subsection{Relativistic effects, characteristic flare shape, and intrinsic timescales}\label{sec:relativistic_effects_intrinsic_time_scales}
In the last section we determined that bright flares in the NIR and X-ray show similar, exponential response functions. This is a generic feature of the light curves that all models of the Sgr~A* emission need to be able to reproduce. In the context of the hot-spot model, the characteristic profile arises from three aspects of the model: (1) the intrinsic variability in the rest frame of the hot spot, that is, the emission generated by the electron plasma; (2) the imprint of the relativistic effects that modulate the light curve as the hot spot moves around the black hole; or (3) a combination of the two.

In \autoref{sec:extreme_cases} we test the two extreme cases: a hot spot of constant intensity orbiting the black hole close to the ISCO and an intrinsic hot-spot variability where the intrinsic rise and fall is exponential with the observed time constants. The study of the two extreme cases illustrates two general conclusions on the observed time-domain characteristics of the variability of Sgr~A* in the hot-spot scenario.

First, the relativistic magnification of a constant source leads to profiles similar to an exponential rise and decay in the observed flux density. They may be asymmetric because at high inclinations a strong lensing magnification dominates the rise and decay of the magnification curve, which is not observed. 

Second, for an intrinsic exponential profile, the relativistic effects generally shorten the intrinsic rise and decay times. This effect depends on the observer inclination: the higher the inclination, the shorter the observed timescale with respect to the intrinsic timescale.

In order to determine the intrinsic timescale, we fitted the observed impulse response function (i.e., the exponential shape) with profiles generated from simulated light curves based on an intrinsic response function with an exponential profile and the relativistic modulation as it would occur for an orbiting hot spot. This allowed us to ``deconvolve'' the observed profile and to derive an estimate of the allowed viewing angles. 
In particular, we constructed the empirical model using the following steps to generate model PCA components that we could compare to the observed shape.

First, the intrinsic emission was modeled as a symmetric double-exponential kernel, with rise and decay time $\tau$.

Second, the relativistic modulation was derived from the \texttt{NERO} ray-tracing code used for \cite{GRAVITYCollaboration2018_orbital, GravityCollaboration2020_orbital}. In order to derive the magnification light curve for each instance of a hot spot, its starting position was drawn from a uniform longitudinal distribution around the black hole. For each instance, the radial separation was derived from a Gamma distribution with a median separation of $\sim8~R_g$ and a minimum separation of $6~R_g$, corresponding to the ISCO of a non-spinning black hole. Finally, the magnification was calculated by assuming an observer inclination angle, $i$, a free parameter.

Third, to account for the noise floor in the PCA of the observed data, we included a free parameter, $\sigma$, for the (Gaussian) noise level in the model flares. Finally, in order to calculate the model response function, we shifted $300$ flares to align their maxima and derived the first PCA component. 

\begin{figure}
    \centering
    \includegraphics[width=0.485\textwidth]{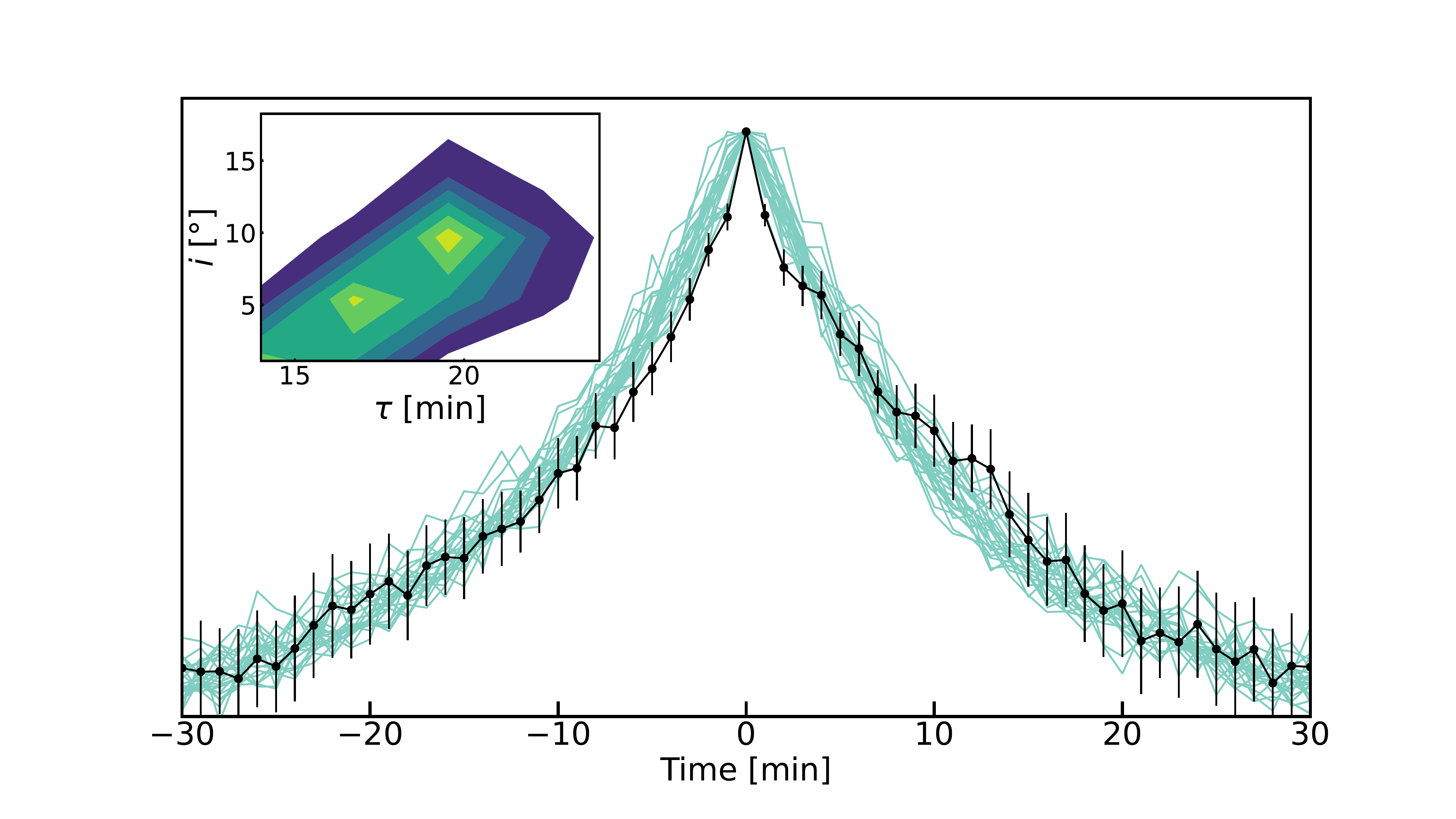}
    \caption{Model PCA components calculated from the posterior samples derived from the flare empirical model (light green) and fit to the observed PCA (black points and line). The errors are determined from the standard deviation of the bootstrap sample (see the text for details). The inset shows the posterior of the observer inclination and intrinsic timescale (see \autoref{sec:relativistic_effects_intrinsic_time_scales} for details and caveats).}
    \label{fig:deconvolved_flare}
\end{figure}

The profile of the model's first PCA component was compared to the observed profile using a $\chi^2$ distance function. We used the \texttt{dynesty} package \citep{Speagle2020_dynesty, Skilling2004_dynesty, Skilling2006_dynesty, Skilling2006_dynesty2} to sample the model parameters ($\tau, i, \sigma$). Because the PCA components are summary statistics of the data and model, $\chi^2$ serves as a distance function rather than a likelihood. The derived posterior samples are thus approximations of the true posterior. \autoref{fig:deconvolved_flare} shows 20 posterior samples together with the posterior contours. The inclination, $i$, was constrained to be below $20\degree$, and the intrinsic rise and fall time can be up to $25\%$ larger than the observed values. Further, the intrinsic timescale is longer at higher inclinations. In other words, the underlying particle acceleration may last longer than flux is observed. We caution that these constraints depend on the choice of the intrinsic kernel. As demonstrated in \autoref{sec:extreme_cases}, the relativistic effects generally lead to an exponential magnification of the light curve, and thus a different kernel (such as a Gaussian) may allow higher inclinations. Our model light curves are idealized with only one flare and no additional Sgr A* variability or correlated noise. Furthermore, the constraints do not account for the effect of the shearing of the hot spot. For instance, in a radial shearing model, as in \citet[shearing due to different orbital speeds within an extended hot spot]{GravityCollaboration2020_orbital}, the relativistic magnification of the light curve, is damped due the integrated relativistic magnification along the sheared hot spot. In such a model, our inclination constraint would be correlated with the size of the hot spot (see the discussion in \cite{GravityCollaboration2020_orbital}, Sects. 5 and 6). Such radial shearing is, however, only one possibility. Further possibilities include shearing due to the distortion of the hot-spot boundary due to (e.g., Rayleigh Taylor) instabilities \citep{Ripperda2022}. We leave the impact of shearing to future work. Nevertheless, the observed timescale may be altered by relativistic effects, which generally leads to higher estimations for the intrinsic timescales at higher inclinations.

\subsection{Phase dispersion minimization analysis of the NIR data}\label{sec:PDM_analysis}

The first component of the PCA can explain a (large) fraction of the variance present in the observed data. To determine whether or not the light curves show evidence of periodicity, we used a phase dispersion minimization (PDM) algorithm implemented in Python (PyPDM). Phase dispersion (PD) is the normalized variance of the data after folding it with a given period. In contrast to standard periodicity analysis tools such as discrete Fourier transforms or Lomb-Scargle periodograms, PDM works well on unevenly sampled data with non-sinusoidal oscillations. Additionally, we expect the periodic signal of individual flares to be ``out of phase'' with one another. Because variance is an additive quantity, it is possible to add the PD curves of individual data segments to accumulate significance for any periodicity that might be present in the data.

To select data segments that represent a flare, we used a similar method as above. However, in order to cover several periods of oscillations, we did not use a threshold for the flux density but instead used one for the fluence (integrated flux density over a given window). This allowed us to identify the segments of the data that are not only bright enough but also remain sufficiently long at elevated flux density levels to potentially show several oscillations. In particular, we used a sliding window of 140~minutes, integrated the flux density over each window (the fluence), and determined the fluence as a function of the midpoint time of each sliding window. Where the (normalized) fluence showed a peak above a threshold of 0.4, we selected a window of $\pm 125$~minutes and applied the PD analysis.

\begin{figure}
    \centering
    \includegraphics[width=0.5\textwidth]{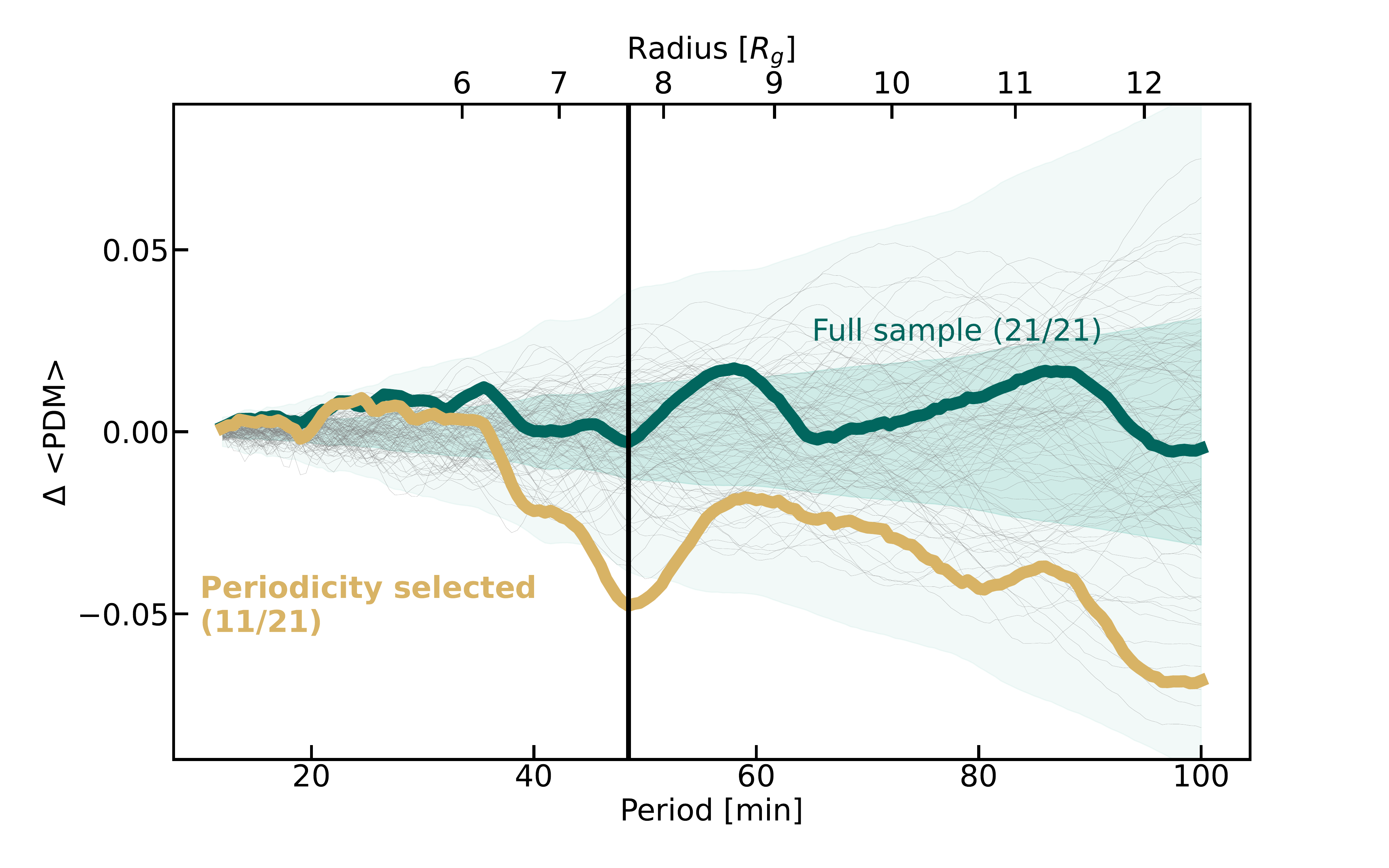}
    \caption{Phase dispersion as a function of period and radius of the hot-spot orbit. The thick green curves show the accumulated PD of all \textit{Spitzer} data segments that lie above the fluence threshold. The gray curves show the accumulated PD of 100 simulated red-noise light curves with auto-correlation properties consistent with the observed \textit{Spitzer} light curves. The green shades show the 1$\sigma$ and 3$\sigma$ contours. The thick yellow curve shows the subset of data segments with the most pronounced periodicity at 50 minutes. This subset contains 11 out of the 21 original data segments.}
    \label{fig:periodicity}
\end{figure}

We found 22 data segments that obey the above criteria. \autoref{fig:periodicity} shows the accumulated PD curves of all these segments and of a hundred simulated red-noise light curves with appropriate auto-correlation properties \citep{2018ApJ...863...15W} after the mean of the simulated PD curves has been subtracted.

Additionally, we folded every data segment with the period of 49~min (the suggested orbital period of the astrometric loops observed with GRAVITY) and fitted a sinus function to the folded data. For 11 out of the 21 data segments, this resulted in an acceptable fit with a large amplitude with respect to the measurement uncertainties. \autoref{fig:periodicity} presents the accumulated PD curve for the 11 selected data segments.

In summary, the \textit{Spitzer} data do not provide independent evidence for periodicity in the range 25-50~minutes. However, as is evident from Fig.~\ref{fig:periodicity}, half of the episodes of increased flux density seem to be consistent with a period in the 25-50~minute range. This is consistent with the expectation of an orbiting hot-spot model in which the flare duration is of similar length as the orbiting period, as suggested by the observations of \cite{GRAVITYCollaboration2018_orbital}. While periodicity cannot be shown with significance, a revolving hot spot as observed by GRAVITY cannot be excluded based on the \textit{Spitzer} light curves, and several of the flares observed with \textit{Spitzer} are in rather good agreement with this a picture. 

\subsection{Relativistic imprints of an orbiting hot spot}\label{sec:relativistic_imprints}
Above, we modeled the observed exponential profile using a toy model for the variability and ignoring the more complex structure in the  variability of Sgr~A* before, during, and after the flares. In this section we extend the simple hot-spot model from the last section and try to fit the NIR flares individually (\autoref{fig:pca_flares}). Essentially, this represents a systematic revisit of the work by \cite{Hamaus2009} and \cite{Karsen2017}. 
Our model consists of three components: an intrinsic variability profile, the relativistic modulation, and a correlated-noise component. The latter serves to describe the observed variability, which we cannot attribute to an instance of a hot spot. Further, there is no a priori (physical) reason for hot spots not to overlap in time, so modeling the light curve with the light curve of a single hot spot may not be adequate.
Explicitly, we modeled flare light curves with the following model:
(1) an intrinsic variability profile, generated by an unspecified process, approximated by a symmetrical Gaussian of amplitude $A$, width $w_0$, and peak time $t_0$; (2) the GR magnification kernel, which is multiplied by the intrinsic flare kernel. The shape of the GR kernel depends on the radial separation from the black hole, $r_0$, the inclination angle of the observer, $i$, and  the phase offset, $\Omega$, in the orbit with respect to the observer;\ (3) a two-parameter red-noise process modeled by a Gaussian process\footnote{To model the Gaussian process, we relied on the python package {George} \citep{george_paper} and used a Mat{\'e}rn $3/2$ kernel function.}, which encapsulates all variability not captured by the flare model.

\begin{figure}
    \centering
    \includegraphics[width=0.485\textwidth]{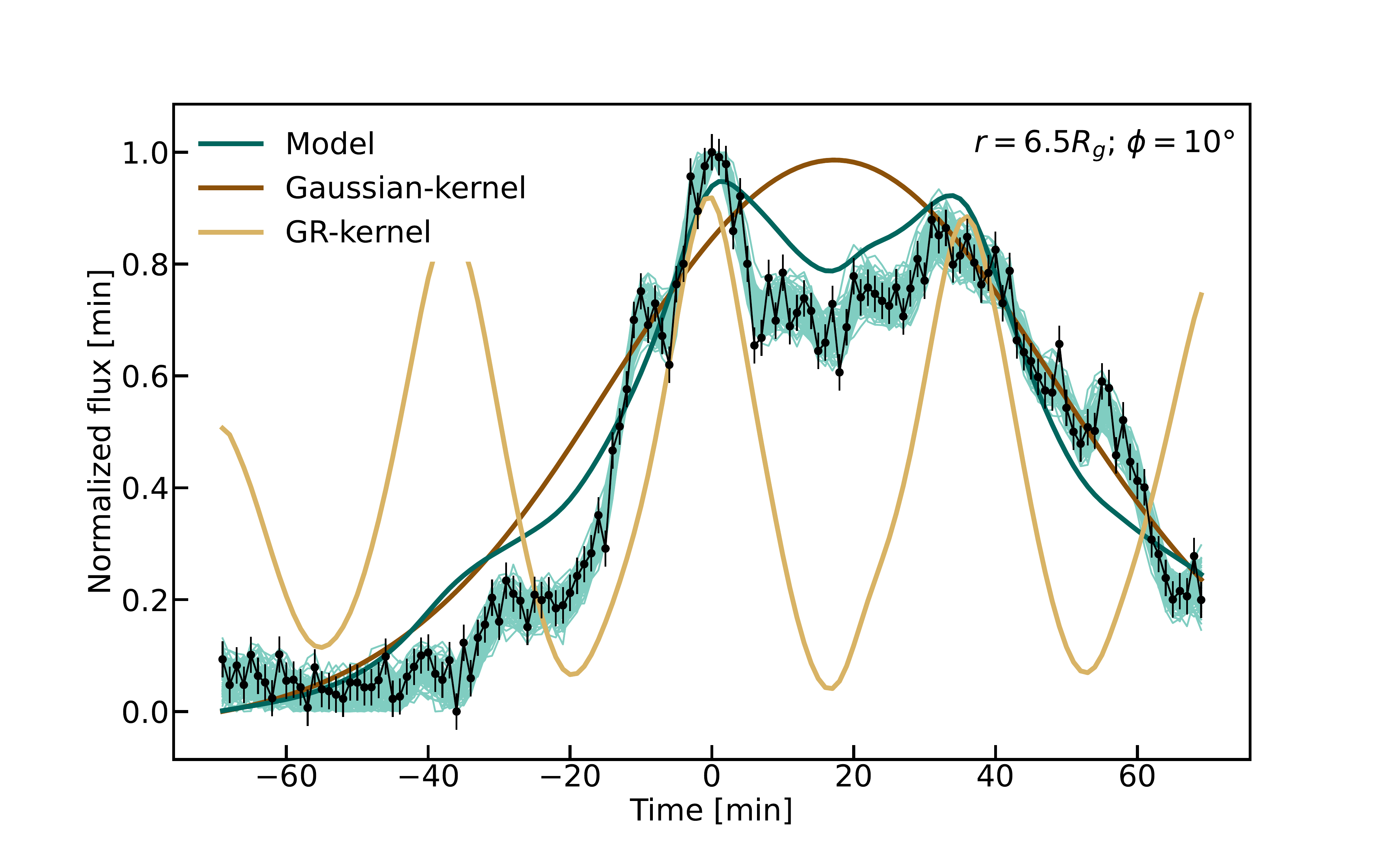}
    \caption{Example of a flare fit with a well-determined width ($t_0$), radial separation ($r_0$), and inclination ($\phi$). We plot the average of $100$ model light curves drawn from the posterior. To make all model components comparable, the fluxes have been normalized. The light yellow line shows the averaged relativistic kernel, dark brown shows the averaged Gaussian flare kernel, dark green shows the composite model without the Gaussian process component, and light green shows the $100$ realizations of the full model.}
    \label{fig:model_fit_example}
\end{figure}

Modeling each \textit{Spitzer} flare as the superposition of a correlated noise process and the light curve of a single hot spot allows the light curve to be fit in a quantitative way. This approach ensures that no spurious features in the light curve are interpreted as a relativistic signal, which would lead to overly tight or even wrong constraints on the relevant parameters ($r_0, i$). We initially fixed the black hole spin to zero (i.e., a non-spinning black hole), as our GR magnification kernel is too coarsely gridded to allow a free spin parameter. In order to compute the posterior for each flare, we again used the software package \texttt{dynesty}. \autoref{fig:model_fit_example} shows an example of a bright flare in which the different model components are well illustrated. We chose a flat prior on the flare orbit radius, $r_0$ ($r_0 \in [6\mathrm{R_g}, 10\mathrm{R_g}]$). This is consistent with the width reported by \cite{GRAVITYCollaboration2018_orbital}, and the lower bound corresponds to the ISCO of a non-spinning black hole. The flare width was constrained to $w_0\in[\SI{5}{\minute}, \SI{90}{\minute}]$, $t_0 \in[-\SI{20}{\minute},\SI{20}{\minute}]$, and the angles were constrained within the respective domains. 

\subsubsection{Fitting results}
\begin{figure*}
    \centering
    \includegraphics[width=0.985\textwidth]{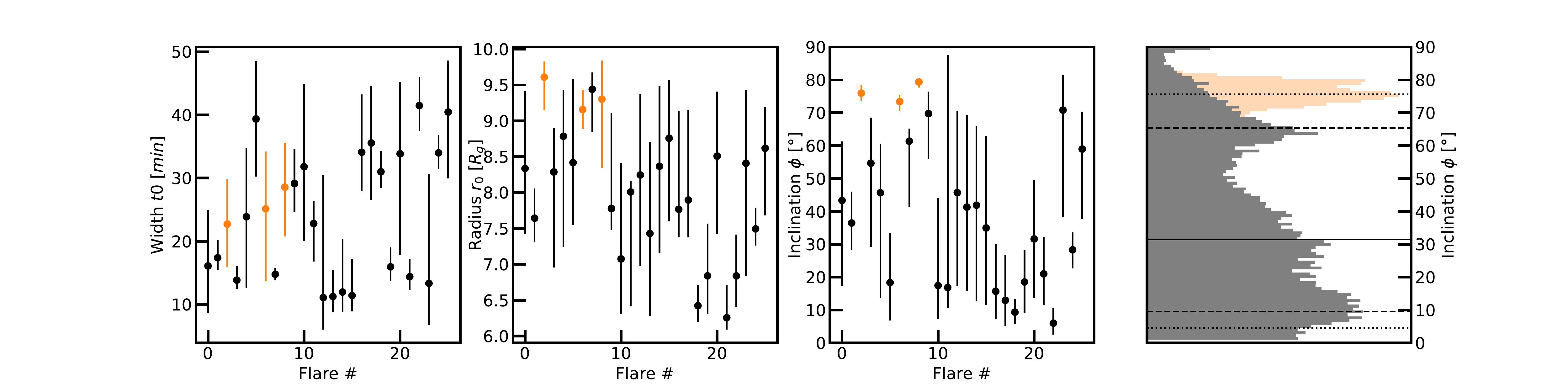}
    \caption{Fit results of the $25$ individual flares. The three left plots show the mean and $1\sigma$ values of the posterior of the respective fits. The three points highlighted in orange show flares that are inconsistent with a low inclination ($i$). The rightmost plot shows the histograms of the stacked posterior samples, both excluding the three high inclination flares (in gray) and  including them (in orange).}
    \label{fig:model_fit_results}
\end{figure*}

\autoref{fig:model_fit_results} shows the fitting results. Three conclusions can be immediately drawn: The flares have typical standard widths of $t_0 \approx \SI{21}{min}$ ($t_{0,\rm{FWHM}}\approx\SI{50}{\min}$). The radial separation is poorly constrained, scattering around almost the entire prior width ($6--10~\mathrm{R_g}$, mean  $\sim 8~\mathrm{R_g}$).
The poor radial constraint can be understood as being due to the absence of strong periodic features in the light curve (see \citealt{Do2009} and our Sect.~\ref{sec:PDM_analysis}). 

Due to the duration of the flares being similar to the orbital period, no multiple revolutions of the same hot spot are observed. The radial separation parameter, $r_0$, would be best constrained by observing multiple orbits and would explain the poor constraint. 
Lastly, the viewing inclination is around \SI{\sim10}{\degree} to \SI{\sim65}{\degree}, with a median inclination of \SI{\sim30}{\degree}. Three flares, however, have well-determined inclinations of around \SI{\sim 80}{\degree} and large separations, forming a group of outliers (highlighted in orange in \autoref{fig:model_fit_results}). 
\noindent Except for these three flares, our results are thus fully consistent with those found by the GRAVITY experiments and the EHT imaging and polarization projects. 

\begin{figure}
\centering
    \begin{subfigure}[b]{0.55\textwidth}
       \includegraphics[width=1\linewidth]{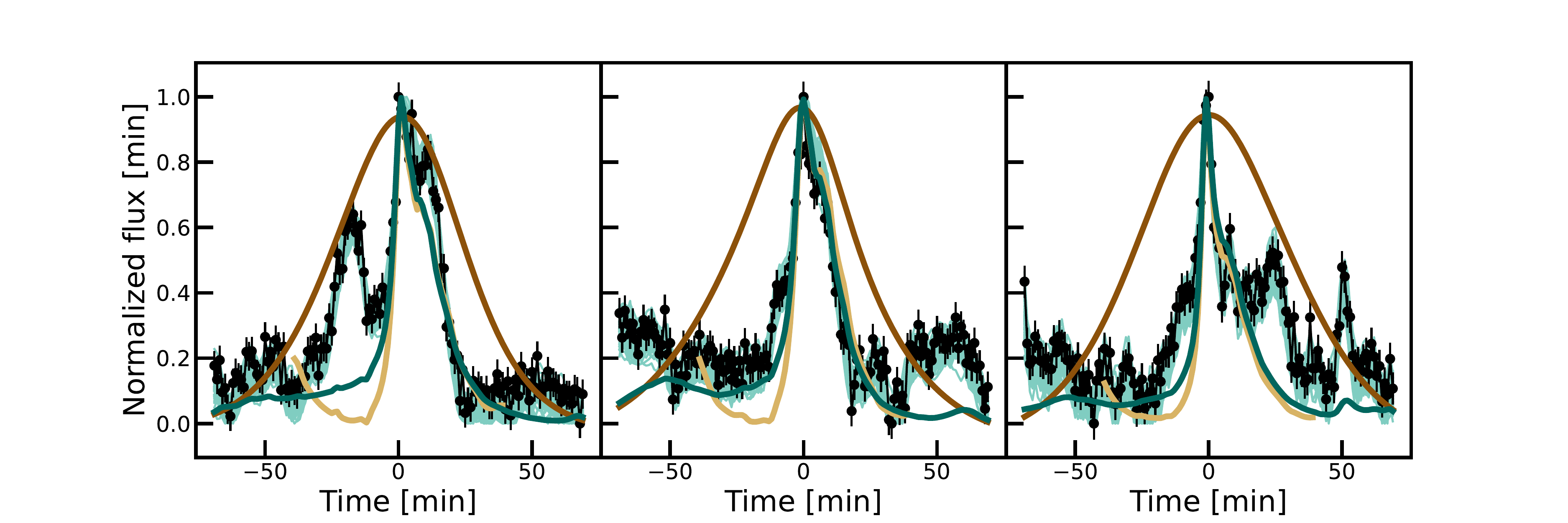}
        \caption{}
        \label{fig:flare_outlier}
    \end{subfigure}
    \centering
    \begin{subfigure}[b]{0.55\textwidth}
       \includegraphics[width=1\linewidth]{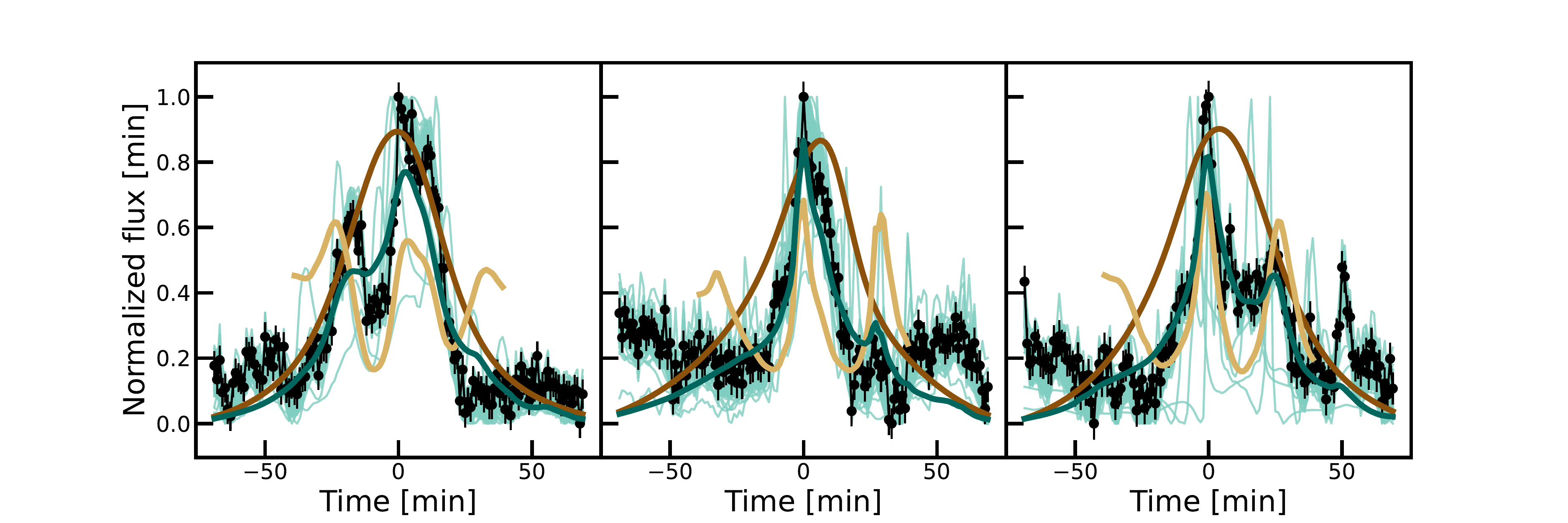}
       \caption{}
       \label{fig:flare_lowrad}
    \end{subfigure}
    \caption{Fit solutions for the three outlying flares. (a) Same as \autoref{fig:model_fit_example}, but for the three flares that require high inclinations and large separations. The secondary peaks are not described by the flare models, but instead are modeled by the Gaussian process component.
(b) Same as panel (a), but showing fit results for a spinning black hole ($a=0.99$), which allows for smaller ISCOs and thus shorter orbital periods ($r_0\sim5.1\mathrm{R_G}$). Here the secondary peaks are described by the flare model.}
\end{figure}

\paragraph{Outlying flares}
\autoref{fig:flare_outlier} shows the model fits for the three outlying flares. All of the flares share a steeply rising flank, requiring a model to have high inclinations and wide orbits. Strikingly, two of the three flares show side peaks, which are too close to one another to be an allowed orbit of a non-spinning black hole (i.e., $r_0 > 6R_g$). In the non-spinning case, these side peaks are not modeled with the flare component of the hot-spot model, but with the Gaussian-process component. This may be interpreted as quiescence flux or flux from a separated flare.
If we allow for the tighter orbits possible around a spinning black hole, for example by setting the prior range to $r_0 \in [5R_g, 6R_g]$, we obtain solutions in which the side peaks are fitted by the flare component. In addition, this leads to lower inclinations, as the observed light curves are inconsistent with the very strong secondary lensing peak that is expected for an edge-on orbit at close separation. The median inclinations for the flares in \autoref{fig:flare_lowrad} are $\SI{30}{\degree}$, $\SI{50}{\degree}$, and $\SI{37}{\degree}$.

\section{Discussion}
\subsection{A NIR-to-X-ray impulse response function}
\subsubsection{Autoregressive representation of a light curve}
One way to model time series is in the framework of the autoregressive (AR) series expansion of the light curve. Any stationary series of data points, $X(n)$,  can be modeled in AR form:
\begin{equation}
    X(n) = \sum_k a_k X(n-k) +  R(n), \label{eqn:ar_process}
\end{equation}
where $R(n)$ is the value of an uncorrelated random process called the innovation at time $n$ \citep[e.g.,][]{Priestley1988}. The coefficients $a_k$ are called the AR coefficients and describe the strength of the correlation of $X(n)$ with the past, $X(n-k)$. Depending on the number, $N$, of coefficients required to describe a given time series, the process is called an AR process of order $N$ - AR($N$). One important realization of the AR process is the Ornstein-Uhlenbeck (O-U) process  \citep{Uhlenbeck1930}, which corresponds to a continuous AR(1) process: a damped random walk. As shown in \autoref{sec:ou_process}, the expectation value for a time series point far away from the process mean follows that of a simple exponential, $\mathbb{E} (x_t)\propto e^{-\theta t}$, consistent with the observed flare shape in the NIR and X-ray. Further, the power spectral density (PSD) slope of an O-U process is $\propto \nu ^ {-2}$, which is consistent with NIR PSD measurements \citep{Do2009, Witzel2018, Witzel2021}. This implies that the PSD slope during flares is indistinguishable from that of the general light curve. This is consistent with the RMS flux relation measurement of \cite{GRAVITYCollaboration2020flux}, which showed an identical RMS flux relation slope during bright flares and faint states\footnote{Five-minute time bins were used.}. As a corollary, this implies that the PSD slope in the X-ray $\propto \nu^{-2}$, at least during flares. 

\subsubsection{Moving average representation of the light curve}
The AR representation of stationary time series is one possible framework. An alternate conceptualization of stationary time series, one that is physically easier to interpret, is the moving average (MA) process:
\begin{equation}
    X(n) = \sum_k c_k R(n-k) + D(n)\label{eqn:ma_proces}
,\end{equation}
where the $c_k$ are the MA coefficients, $D(n)$ is a deterministic process typically set to zero, and $R(\cdot)$ is the same innovation (that is, identical random seeds) as in \autoref{eqn:ar_process}. In the MA representation, the coefficients $c_k$ are typically interpreted as the impulse response of the system. The MA and AR presentations of light curve are convolutional inverses of one another: formally, an AR(1) process corresponds to an infinite MA process with an exponentially decaying pulse shape \citep{Scargle1981}, as observed. \cite{Scargle2020} demonstrated that, depending on the sparsity of the innovation, the intrinsic impulse response of the light-curve-generating process matches the profile of individual, observed flares. In our context, we argue that the PCA is able to identify one set of $c_k$ that describes the system response of Sgr~A* well, with very limited biases, even in the presence of high correlated noise (\autoref{sec:pca_shape} and \autoref{sec:pca_noise_bias}).

Several authors have argued against such an event-like nature for Sgr~A* \citep{Do2009, Meyer2014}, reasoning that the correlated red-noise behavior observed for Sgr~A* is inconsistent with or disfavors an event-like nature of the flux outburst, typically employing AR methods to quantify the variability behavior. However, neither representation implies a physical system, and they can be converted from one to another \citep[see, for instance, the extensive discussion in][]{Scargle2020}. Thus, the presence of correlated (red-noise) flux does not imply that the flux-generating process {must} correspond to a non-event-like physical process; conversely, the presence of flares in the light curve does not imply an event-like process. Of course, both time series expansions can be applied to the observed data, with the AR process posing a more intuitive basis for a continuous process and the MA process more intuitive for event-like flares, but they are otherwise purely mathematical concepts. In essence, auxiliary evidence is required to argue for a continuous or event-like process, and the argumentation based on the successful description of the data with one or the other expansion is false. 

For Sgr~A*, several observational arguments favor an event-like nature of its flares: First, the flux distribution can be modeled by a two-component distribution consisting of a log-normal with a power-law tail. Also, other non-lognormal, log-right-skewed distributions match the data \citep{Dodds-Eden2009, Meyer2014, Do2019, GRAVITYCollaboration2020flux}. Second, the spectral slope during high flux states is positive ($\nu L_ \nu \propto \nu^{\sim +0.5}$), while for fainter flux states it decreases, indicating a transition from flaring to quiescence \citep{Eisenhauer2005, Krabbe2006, Gillessen2006, Dodds-Eden2010, Hornstein2007, Eckart2009, Ponti2017, Witzel2018, GravityCollaboration2021_xrayflare, Boyce2022}.
 
Third, the SED of NIR--X-ray flares can be modeled with a synchrotron or synchrotron self-Compton sphere localized in the accretion flow \citep{Dodds-Eden2011, Ponti2017, Witzel2021, GravityCollaboration2021_xrayflare, Boyce2022}. Fourth, the flare of Sgr~A* shows signatures in linear polarization (Q--U loops) in NIR and submillimeter observations, consistent with a localized hot spot in the accretion flow \citep{Marrone2006, GravityCollaboration2020_polarization_ale, Wielgus2022}. Finally, the centroid of light emission showed a clockwise loop during three NIR flares observed with the GRAVITY interferometer \citep{GRAVITYCollaboration2018_orbital}, again consistent with the hot-spot picture. 

Furthermore, NIR and X-ray observations are notoriously hard to model in modern GRMHD simulations. To date, no such simulation is capable of capturing all observational aspects, and GRMHD simulations that capture the submillimeter emission well cannot model the nonthermal emission observed in the NIR and X-ray \citep[e.g.,][]{eht_paper_V}. The nonthermal emission is typically attributed to event-like processes such as magnetic reconnection, turbulent heating, or shocks in the accretion flow \citep{Ripperda2020, Dexter2014, Dexter2020_model, Ball2016}. 

We argue that the observed exponential rise and decay profile with a characteristic time of $\tau \sim \SI{15}{\minute}$ serves as a quantifiable observable to constrain the flare emission mechanism. However, care must be taken to account for relativistic effects because the intrinsic timescales are longer than the observed one. In contrast to other measures of the variability, such as the power spectrum, the RMS flux relation, the flux distribution, or simultaneous flux measurements of the flare SED, it represents a high-S/N, event-averaged, and spectrum-averaged measurement of the variability characteristics.

The characteristic shape manifested in the total intensity measurements of flares may also be present in the polarization light curves. The Q-U loops observed by \cite{GRAVITYCollaboration2018_orbital, GravityCollaboration2020_polarization_ale} and \cite{Wielgus2022} may be a first indication that such a polarization response indeed exits. Observations of a similar impulse response in polarization may therefore allow one to disentangle between relativistic effects and the properties of the intrinsic emission.

\subsection{Relativistic modeling of flares}
In addition to the exponential shape common to all flares, many of the flares show substructure. This substructure is expected in the context of the  orbiting hot-spot model. In this model, the intrinsic flux is modulated by the relativistic effects of the hot spot traveling close to the black hole. Subsection \ref{sec:relativistic_imprints} shows that the observed flares are consistent with a black hole viewed at low inclination, with a typical flare width of $\SI{\sim25}{\minute}$. Based on the observed flares we can rule out edge-on orientations for all but three flares. Two of the three outlying flares show characteristic double-peaked substructure, and all can be modeled with a lower inclination if one allows for a spinning black hole ($a=0.99$). 

\section{Summary and conclusions}
In this paper we have analyzed the $25$ high flux states of Sgr~A* in eight \textit{Spitzer} observations of roughly $24$ hours each. In addition, we extracted $24$ high flux states from almost $1500$ hours of \textit{Chandra} X-ray observations. We defined a window of $\SI{\pm70}{\minute}$ around the peak of each flare, on which we centered, stacked, and normalized each data segment. This reveals a characteristic symmetric and exponential response function. A PCA decomposition reveals that $\sim50\%$ and $75\%$ of the variance in the NIR and X-ray flares can be explained by this characteristic profile, which may indicate that emission from secondary processes (for instance, flux from synchrotron cooled electrons) contributes to the observed NIR flares. The shape is very well fit by a symmetrical exponential rise and decay with a rise and decay time of $\tau\SI{\sim15}{\minute}$. No significant asymmetry in the flare shape could be shown.

In the context of the hot-spot model, we consider it likely that this profile is generated by the superposition of an intrinsic response function with relativistic effects. However, under the assumption of an intrinsic exponential shape, relativistic effects shorten the observed timescales depending on the hot-spot parameters. Importantly, we find the intrinsic symmetrical shape inconsistent with hot spots viewed at high inclinations. 

Lastly, modeling the light curve of the $25$ \textit{Spitzer} flares with a generic hot-spot model, we find that all but three flares are consistently modeled by a black hole--hot-spot system viewed under a low inclination ($\phi_{\rm{median}} \sim 30 \degree$). Inclinations higher than $75\degree$ are disfavored ($3\sigma$). Three flares require higher inclinations if only non-spinning black holes are considered. If one allows for nonzero black hole spin, with correspondingly tighter ISCOs, all flares are consistent with being viewed under low inclinations. 

\begin{acknowledgements}
     This work was supported in part by the Deutsche Forschungsgemeinschaft (DFG) via the Cologne Bonn Graduate School (BCGS), the Max Planck Society through the International Max Planck Research School (IMPRS) for Astronomy and Astrophysics, as well as special funds through the University of Cologne and the DFG CRC956 ‘Conditions and Impact of Star Formation’ under project A2.
\end{acknowledgements}

\bibliography{bib}

\begin{thebibliography}{94}
\expandafter\ifx\csname natexlab\endcsname\relax\def\natexlab#1{#1}\fi

\bibitem[{{Ambikasaran} {et~al.}(2014){Ambikasaran}, {Foreman-Mackey},
  {Greengard}, {Hogg}, \& {O'Neil}}]{george_paper}
{Ambikasaran}, S., {Foreman-Mackey}, D., {Greengard}, L., {Hogg}, D.~W., \&
  {O'Neil}, M. 2014

\bibitem[{{Ball} {et~al.}(2016){Ball}, {{\"O}zel}, {Psaltis}, \&
  {Chan}}]{Ball2016}
{Ball}, D., {{\"O}zel}, F., {Psaltis}, D., \& {Chan}, C.-k. 2016, \apj, 826, 77

\bibitem[{{Barri{\`e}re} {et~al.}(2014){Barri{\`e}re}, {Tomsick}, {Baganoff},
  {Boggs}, {Christensen}, {Craig}, {Dexter}, {Grefenstette}, {Hailey},
  {Harrison}, {Madsen}, {Mori}, {Stern}, {Zhang}, {Zhang}, \&
  {Zoglauer}}]{Barriere2014}
{Barri{\`e}re}, N.~M., {Tomsick}, J.~A., {Baganoff}, F.~K., {et~al.} 2014,
  \apj, 786, 46

\bibitem[{{Bower} {et~al.}(2018){Bower}, {Broderick}, {Dexter}, {Doeleman},
  {Falcke}, {Fish}, {Johnson}, {Marrone}, {Moran}, {Moscibrodzka}, {Peck},
  {Plambeck}, \& {Rao}}]{Bower2018}
{Bower}, G.~C., {Broderick}, A., {Dexter}, J., {et~al.} 2018, \apj, 868, 101

\bibitem[{Bower {et~al.}(2019)Bower, Dexter, Asada, Brinkerink, Falcke, Ho,
  Inoue, Markoff, Marrone, Matsushita, Moscibrodzka, Nakamura, Peck, \&
  Rao}]{Bower2019}
Bower, G.~C., Dexter, J., Asada, K., {et~al.} 2019, \apj, 881, L2

\bibitem[{{Bower} {et~al.}(2015){Bower}, {Markoff}, {Dexter}, {Gurwell},
  {Moran}, {Brunthaler}, {Falcke}, {Fragile}, {Maitra}, {Marrone}, {Peck},
  {Rushton}, \& {Wright}}]{Bower2015}
{Bower}, G.~C., {Markoff}, S., {Dexter}, J., {et~al.} 2015, \apj, 802, 69

\bibitem[{{Boyce} {et~al.}(2022){Boyce}, {Haggard}, {Witzel}, {Fellenberg},
  {Willner}, {Becklin}, {Do}, {Eckart}, {Fazio}, {Gurwell}, {Hora}, {Markoff},
  {Morris}, {Neilsen}, {Nowak}, {Smith}, \& {Zhang}}]{Boyce2022}
{Boyce}, H., {Haggard}, D., {Witzel}, G., {et~al.} 2022, \apj, 931, 7

\bibitem[{{Boyce} {et~al.}(2021){Boyce}, {Haggard}, {Witzel}, {Willner},
  {Neilsen}, {Hora}, {Markoff}, {Ponti}, {Baganoff}, {Becklin}, {Fazio},
  {Lowrance}, {Morris}, \& {Smith}}]{Boyce2021}
{Boyce}, H., {Haggard}, D., {Witzel}, G., {et~al.} 2021, \apj, 912, 168

\bibitem[{{Boyce} {et~al.}(2019){Boyce}, {Haggard}, {Witzel}, {Willner},
  {Neilsen}, {Hora}, {Markoff}, {Ponti}, {Baganoff}, {Becklin}, {Fazio},
  {Lowrance}, {Morris}, \& {Smith}}]{Boyce2019}
{Boyce}, H., {Haggard}, D., {Witzel}, G., {et~al.} 2019, \apj, 871, 161

\bibitem[{{Bransgrove} {et~al.}(2021){Bransgrove}, {Ripperda}, \&
  {Philippov}}]{Bransgrove2021}
{Bransgrove}, A., {Ripperda}, B., \& {Philippov}, A. 2021, \prl, 127, 055101

\bibitem[{{Brinkerink} {et~al.}(2015){Brinkerink}, {Falcke}, {Law}, {Barkats},
  {Bower}, {Brunthaler}, {Gammie}, {Impellizzeri}, {Markoff}, {Menten},
  {Moscibrodzka}, {Peck}, {Rushton}, {Schaaf}, \& {Wright}}]{Brinkerink2015}
{Brinkerink}, C.~D., {Falcke}, H., {Law}, C.~J., {et~al.} 2015, \aap, 576, A41

\bibitem[{{Brinkerink} {et~al.}(2016){Brinkerink}, {M{\"u}ller}, {Falcke},
  {Bower}, {Krichbaum}, {Castillo}, {Deller}, {Doeleman}, {Fraga-Encinas},
  {Goddi}, {Hern{\'a}ndez-G{\'o}mez}, {Hughes}, {Kramer}, {L{\'e}on-Tavares},
  {Loinard}, {Monta{\~n}a}, {Mo{\'s}cibrodzka}, {Ortiz-Le{\'o}n},
  {Sanchez-Arguelles}, {Tilanus}, {Wilson}, \& {Zensus}}]{Brinkerink2016}
{Brinkerink}, C.~D., {M{\"u}ller}, C., {Falcke}, H., {et~al.} 2016, \mnras,
  462, 1382

\bibitem[{{Broderick} \& {Loeb}(2006)}]{Broderick2006_hotspot}
{Broderick}, A.~E. \& {Loeb}, A. 2006, \mnras, 367, 905

\bibitem[{{Chan} {et~al.}(2015){Chan}, {Psaltis}, {{\"O}zel}, {Medeiros},
  {Marrone}, {Sa{\c d}owski}, \& {Narayan}}]{Chan2015}
{Chan}, C.-k., {Psaltis}, D., {{\"O}zel}, F., {et~al.} 2015, \apj, 812, 103

\bibitem[{{Chatterjee} {et~al.}(2021){Chatterjee}, {Markoff}, {Neilsen},
  {Younsi}, {Witzel}, {Tchekhovskoy}, {Yoon}, {Ingram}, {van der Klis},
  {Boyce}, {Do}, {Haggard}, \& {Nowak}}]{Chatterjee2020}
{Chatterjee}, K., {Markoff}, S., {Neilsen}, J., {et~al.} 2021, \mnras, 507,
  5281

\bibitem[{{Chen} {et~al.}(2018){Chen}, {Yuan}, \& {Yang}}]{Chen2018}
{Chen}, A.~Y., {Yuan}, Y., \& {Yang}, H. 2018, \apjl, 863, L31

\bibitem[{{Comisso} {et~al.}(2020){Comisso}, {Sobacchi}, \&
  {Sironi}}]{Comisso2020}
{Comisso}, L., {Sobacchi}, E., \& {Sironi}, L. 2020, \apjl, 895, L40

\bibitem[{{Crinquand} {et~al.}(2022){Crinquand}, {Cerutti}, {Dubus}, {Parfrey},
  \& {Philippov}}]{Crinquand2022}
{Crinquand}, B., {Cerutti}, B., {Dubus}, G., {Parfrey}, K., \& {Philippov}, A.
  2022, \prl, 129, 205101

\bibitem[{{Crinquand} {et~al.}(2020){Crinquand}, {Cerutti}, {Philippov},
  {Parfrey}, \& {Dubus}}]{Crinquand2020}
{Crinquand}, B., {Cerutti}, B., {Philippov}, A., {Parfrey}, K., \& {Dubus}, G.
  2020, \prl, 124, 145101

\bibitem[{{Davelaar} {et~al.}(2018){Davelaar}, {Mo{\'s}cibrodzka}, {Bronzwaer},
  \& {Falcke}}]{Davelaar2018}
{Davelaar}, J., {Mo{\'s}cibrodzka}, M., {Bronzwaer}, T., \& {Falcke}, H. 2018,
  \aap, 612, A34

\bibitem[{{Dexter} {et~al.}(2010){Dexter}, {Agol}, {Fragile}, \&
  {McKinney}}]{Dexter2010}
{Dexter}, J., {Agol}, E., {Fragile}, P.~C., \& {McKinney}, J.~C. 2010, \apj,
  717, 1092

\bibitem[{{Dexter} {et~al.}(2020{\natexlab{a}}){Dexter}, {Jim{\'e}nez-Rosales},
  {Ressler}, {Tchekhovskoy}, {Baub{\"o}ck}, {de Zeeuw}, {Eisenhauer}, {von
  Fellenberg}, {Gao}, {Genzel}, {Gillessen}, {Habibi}, {Ott}, {Stadler},
  {Straub}, \& {Widmann}}]{Dexter2020_model}
{Dexter}, J., {Jim{\'e}nez-Rosales}, A., {Ressler}, S.~M., {et~al.}
  2020{\natexlab{a}}, \mnras, 494, 4168

\bibitem[{{Dexter} {et~al.}(2014){Dexter}, {Kelly}, {Bower}, {Marrone},
  {Stone}, \& {Plambeck}}]{Dexter2014}
{Dexter}, J., {Kelly}, B., {Bower}, G.~C., {et~al.} 2014, \mnras, 442, 2797

\bibitem[{{Dexter} {et~al.}(2020{\natexlab{b}}){Dexter}, {Tchekhovskoy},
  {Jim{\'e}nez-Rosales}, {Ressler}, {Baub{\"o}ck}, {Dallilar}, {de Zeeuw},
  {Eisenhauer}, {von Fellenberg}, {Gao}, {Genzel}, {Gillessen}, {Habibi},
  {Ott}, {Stadler}, {Straub}, \& {Widmann}}]{Dexter2020_flare}
{Dexter}, J., {Tchekhovskoy}, A., {Jim{\'e}nez-Rosales}, A., {et~al.}
  2020{\natexlab{b}}, \mnras, 497, 4999

\bibitem[{{Do} {et~al.}(2009){Do}, {Ghez}, {Morris}, {Yelda}, {Meyer}, {Lu},
  {Hornstein}, \& {Matthews}}]{Do2009}
{Do}, T., {Ghez}, A.~M., {Morris}, M.~R., {et~al.} 2009, \apj, 691, 1021

\bibitem[{Do {et~al.}(2019)Do, Witzel, Gautam, Chen, Ghez, Morris, Becklin,
  Ciurlo, Hosek, Martinez, Matthews, Sakai, \& Sch{\"{o}}del}]{Do2019}
Do, T., Witzel, G., Gautam, A.~K., {et~al.} 2019, \apj, 882, L27

\bibitem[{{Dodds-Eden} {et~al.}(2011){Dodds-Eden}, {Gillessen}, {Fritz},
  {Eisenhauer}, {Trippe}, {Genzel}, {Ott}, {Bartko}, {Pfuhl}, {Bower},
  {Goldwurm}, {Porquet}, {Trap}, \& {Yusef-Zadeh}}]{Dodds-Eden2011}
{Dodds-Eden}, K., {Gillessen}, S., {Fritz}, T.~K., {et~al.} 2011, \apj, 728, 37

\bibitem[{{Dodds-Eden} {et~al.}(2009){Dodds-Eden}, {Porquet}, {Trap},
  {Quataert}, {Haubois}, {Gillessen}, {Grosso}, {Pantin}, {Falcke}, {Rouan},
  {Genzel}, {Hasinger}, {Goldwurm}, {Yusef-Zadeh}, {Clenet}, {Trippe},
  {Lagage}, {Bartko}, {Eisenhauer}, {Ott}, {Paumard}, {Perrin}, {Yuan},
  {Fritz}, \& {Mascetti}}]{Dodds-Eden2009}
{Dodds-Eden}, K., {Porquet}, D., {Trap}, G., {et~al.} 2009, \apj, 698, 676

\bibitem[{Dodds-Eden {et~al.}(2010)Dodds-Eden, Sharma, Quataert, Genzel,
  Gillessen, Eisenhauer, \& Porquet}]{Dodds-Eden2010}
Dodds-Eden, K., Sharma, P., Quataert, E., {et~al.} 2010, \apj, 725, 450

\bibitem[{{Eatough} {et~al.}(2013){Eatough}, {Falcke}, {Karuppusamy}, {Lee},
  {Champion}, {Keane}, {Desvignes}, {Schnitzeler}, {Spitler}, {Kramer},
  {Klein}, {Bassa}, {Bower}, {Brunthaler}, {Cognard}, {Deller}, {Demorest},
  {Freire}, {Kraus}, {Lyne}, {Noutsos}, {Stappers}, \& {Wex}}]{Eatough2013}
{Eatough}, R.~P., {Falcke}, H., {Karuppusamy}, R., {et~al.} 2013, \nat, 501,
  391

\bibitem[{{Eckart} {et~al.}(2009){Eckart}, {Baganoff}, {Morris}, {Kunneriath},
  {Zamaninasab}, {Witzel}, {Sch{\"o}del}, {Garc{\'\i}a-Mar{\'\i}n}, {Meyer},
  {Bower}, {Marrone}, {Bautz}, {Brandt}, {Garmire}, {Ricker}, {Straubmeier},
  {Roberts}, {Muzic}, {Mauerhan}, \& {Zensus}}]{Eckart2009}
{Eckart}, A., {Baganoff}, F.~K., {Morris}, M.~R., {et~al.} 2009, \aap, 500, 935

\bibitem[{{Eckart} {et~al.}(2012){Eckart}, {Garc{\'{\i}}a-Mar{\'{\i}}n},
  {Vogel}, {Teuben}, {Morris}, {Baganoff}, {Dexter}, {Sch{\"o}del}, {Witzel},
  {Valencia-S.}, {Karas}, {Kunneriath}, {Straubmeier}, {Moser}, {Sabha},
  {Buchholz}, {Zamaninasab}, {Mu{\v z}i{\'c}}, {Moultaka}, \&
  {Zensus}}]{Eckart2012}
{Eckart}, A., {Garc{\'{\i}}a-Mar{\'{\i}}n}, M., {Vogel}, S.~N., {et~al.} 2012,
  \aap, 537, A52

\bibitem[{{Eisenhauer} {et~al.}(2005){Eisenhauer}, {Genzel}, {Alexander},
  {Abuter}, {Paumard}, {Ott}, {Gilbert}, {Gillessen}, {Horrobin}, {Trippe},
  {Bonnet}, {Dumas}, {Hubin}, {Kaufer}, {Kissler-Patig}, {Monnet},
  {Str{\"o}bele}, {Szeifert}, {Eckart}, {Sch{\"o}del}, \&
  {Zucker}}]{Eisenhauer2005}
{Eisenhauer}, F., {Genzel}, R., {Alexander}, T., {et~al.} 2005, \apj, 628, 246

\bibitem[{{Event Horizon Telescope Collaboration}
  {et~al.}(2022{\natexlab{a}}){Event Horizon Telescope Collaboration},
  {Akiyama}, {Alberdi}, {Alef}, {Algaba}, {Anantua}, {Asada}, {Azulay}, {Bach},
  {Baczko}, {Ball}, {Balokovi{\'c}}, {Barrett}, {Baub{\"o}ck}, {Benson},
  {Bintley}, {Blackburn}, {Blundell}, {Bouman}, {Bower}, {Boyce}, {Bremer},
  {Brinkerink}, {Brissenden}, {Britzen}, {Broderick}, {Broguiere}, {Bronzwaer},
  {Bustamante}, {Byun}, {Carlstrom}, {Ceccobello}, {Chael}, {Chan},
  {Chatterjee}, {Chatterjee}, {Chen}, {Chen}, {Cheng}, {Cho}, {Christian},
  {Conroy}, {Conway}, {Cordes}, {Crawford}, {Crew}, {Cruz-Osorio}, {Cui},
  {Davelaar}, {De Laurentis}, {Deane}, {Dempsey}, {Desvignes}, {Dexter},
  {Dhruv}, {Doeleman}, {Dougal}, {Dzib}, {Eatough}, {Emami}, {Falcke}, {Farah},
  {Fish}, {Fomalont}, {Ford}, {Fraga-Encinas}, {Freeman}, {Friberg}, {Fromm},
  {Fuentes}, {Galison}, {Gammie}, {Garc{\'\i}a}, {Gentaz}, {Georgiev}, {Goddi},
  {Gold}, {G{\'o}mez-Ruiz}, {G{\'o}mez}, {Gu}, {Gurwell}, {Hada}, {Haggard},
  {Haworth}, {Hecht}, {Hesper}, {Heumann}, {Ho}, {Ho}, {Honma}, {Huang},
  {Huang}, {Hughes}, {Ikeda}, {Impellizzeri}, {Inoue}, {Issaoun}, {James},
  {Jannuzi}, {Janssen}, {Jeter}, {Jiang}, {Jim{\'e}nez-Rosales}, {Johnson},
  {Jorstad}, {Joshi}, {Jung}, {Karami}, {Karuppusamy}, {Kawashima}, {Keating},
  {Kettenis}, {Kim}, {Kim}, {Kim}, {Kim}, {Kino}, {Koay}, {Kocherlakota},
  {Kofuji}, {Koch}, {Koyama}, {Kramer}, {Kramer}, {Krichbaum}, {Kuo}, {La
  Bella}, {Lauer}, {Lee}, {Lee}, {Leung}, {Levis}, {Li}, {Lico}, {Lindahl},
  {Lindqvist}, {Lisakov}, {Liu}, {Liu}, {Liuzzo}, {Lo}, {Lobanov}, {Loinard},
  {Lonsdale}, {Lu}, {Mao}, {Marchili}, {Markoff}, {Marrone}, {Marscher},
  {Mart{\'\i}-Vidal}, {Matsushita}, {Matthews}, {Medeiros}, {Menten},
  {Michalik}, {Mizuno}, {Mizuno}, {Moran}, {Moriyama}, {Moscibrodzka},
  {M{\"u}ller}, {Mus}, {Musoke}, {Myserlis}, {Nadolski}, {Nagai}, {Nagar},
  {Nakamura}, {Narayan}, {Narayanan}, {Natarajan}, {Nathanail}, {Navarro
  Fuentes}, {Neilsen}, {Neri}, {Ni}, {Noutsos}, {Nowak}, {Oh}, {Okino},
  {Olivares}, {Ortiz-Le{\'o}n}, {Oyama}, {{\"O}zel}, {Palumbo}, {Paraschos},
  {Park}, {Parsons}, {Patel}, {Pen}, {Pesce}, {Pi{\'e}tu}, {Plambeck},
  {PopStefanija}, {Porth}, {P{\"o}tzl}, {Prather}, {Preciado-L{\'o}pez},
  {Psaltis}, {Pu}, {Ramakrishnan}, {Rao}, {Rawlings}, {Raymond}, {Rezzolla},
  {Ricarte}, {Ripperda}, {Roelofs}, {Rogers}, {Ros}, {Romero-Ca{\~n}izales},
  {Roshanineshat}, {Rottmann}, {Roy}, {Ruiz}, {Ruszczyk}, {Rygl},
  {S{\'a}nchez}, {S{\'a}nchez-Arg{\"u}elles}, {S{\'a}nchez-Portal}, {Sasada},
  {Satapathy}, {Savolainen}, {Schloerb}, {Schonfeld}, {Schuster}, {Shao},
  {Shen}, {Small}, {Sohn}, {SooHoo}, {Souccar}, {Sun}, {Tazaki}, {Tetarenko},
  {Tiede}, {Tilanus}, {Titus}, {Torne}, {Traianou}, {Trent}, {Trippe}, {Turk},
  {van Bemmel}, {van Langevelde}, {van Rossum}, {Vos}, {Wagner},
  {Ward-Thompson}, {Wardle}, {Weintroub}, {Wex}, {Wharton}, {Wielgus}, {Wiik},
  {Witzel}, {Wondrak}, {Wong}, {Wu}, {Yamaguchi}, {Yoon}, {Young}, {Young},
  {Younsi}, {Yuan}, {Yuan}, {Zensus}, {Zhang}, {Zhao}, {Zhao}, {Agurto},
  {Allardi}, {Amestica}, {Araneda}, {Arriagada}, {Berghuis}, {Bertarini},
  {Berthold}, {Blanchard}, {Brown}, {C{\'a}rdenas}, {Cantzler}, {Caro},
  {Castillo-Dom{\'\i}nguez}, {Chan}, {Chang}, {Chang}, {Chang}, {Chang},
  {Chen}, {Chilson}, {Chuter}, {Ciechanowicz}, {Colin-Beltran}, {Coulson},
  {Crowley}, {Degenaar}, {Dornbusch}, {Dur{\'a}n}, {Everett}, {Faber},
  {Forster}, {Fuchs}, {Gale}, {Geertsema}, {Gonz{\'a}lez}, {Graham}, {Gueth},
  {Halverson}, {Han}, {Han}, {Hasegawa}, {Hern{\'a}ndez-Rebollar}, {Herrera},
  {Herrero-Illana}, {Heyminck}, {Hirota}, {Hoge}, {Hostler Schimpf}, {Howie},
  {Huang}, {Jiang}, {Jinchi}, {John}, {Kimura}, {Klein}, {Kubo}, {Kuroda},
  {Kwon}, {Lacasse}, {Laing}, {Leitch}, {Li}, {Liu}, {Liu}, {Lin}, {Lu},
  {Mac-Auliffe}, {Martin-Cocher}, {Matulonis}, {Maute}, {Messias},
  {Meyer-Zhao}, {Monta{\~n}a}, {Montenegro-Montes}, {Montgomerie}, {Moreno
  Nolasco}, {Muders}, {Nishioka}, {Norton}, {Nystrom}, {Ogawa}, {Olivares},
  {Oshiro}, {P{\'e}rez-Beaupuits}, {Parra}, {Phillips}, {Poirier}, {Pradel},
  {Qiu}, {Raffin}, {Rahlin}, {Ram{\'\i}rez}, {Ressler}, {Reynolds},
  {Rodr{\'\i}guez-Montoya}, {Saez-Madain}, {Santana}, {Shaw}, {Shirkey},
  {Silva}, {Snow}, {Sousa}, {Sridharan}, {Stahm}, {Stark}, {Test},
  {Torstensson}, {Venegas}, {Walther}, {Wei}, {White}, {Wieching}, {Wijnands},
  {Wouterloot}, {Yu}, {Yu}, {Zeballos}, \& {EHT Collaboration}}]{eht_sgra_I}
{Event Horizon Telescope Collaboration}, {Akiyama}, K., {Alberdi}, A., {et~al.}
  2022{\natexlab{a}}, \apjl, 930, L12

\bibitem[{{Event Horizon Telescope Collaboration}
  {et~al.}(2022{\natexlab{b}}){Event Horizon Telescope Collaboration},
  {Akiyama}, {Alberdi}, {Alef}, {Algaba}, {Anantua}, {Asada}, {Azulay}, {Bach},
  {Baczko}, {Ball}, {Balokovi{\'c}}, {Barrett}, {Baub{\"o}ck}, {Benson},
  {Bintley}, {Blackburn}, {Blundell}, {Bouman}, {Bower}, {Boyce}, {Bremer},
  {Brinkerink}, {Brissenden}, {Britzen}, {Broderick}, {Broguiere}, {Bronzwaer},
  {Bustamante}, {Byun}, {Carlstrom}, {Ceccobello}, {Chael}, {Chan},
  {Chatterjee}, {Chatterjee}, {Chen}, {Chen}, {Cheng}, {Cho}, {Christian},
  {Conroy}, {Conway}, {Cordes}, {Crawford}, {Crew}, {Cruz-Osorio}, {Cui},
  {Davelaar}, {De Laurentis}, {Deane}, {Dempsey}, {Desvignes}, {Dexter},
  {Dhruv}, {Doeleman}, {Dougal}, {Dzib}, {Eatough}, {Emami}, {Falcke}, {Farah},
  {Fish}, {Fomalont}, {Ford}, {Fraga-Encinas}, {Freeman}, {Friberg}, {Fromm},
  {Fuentes}, {Galison}, {Gammie}, {Garc{\'\i}a}, {Gentaz}, {Georgiev}, {Goddi},
  {Gold}, {G{\'o}mez-Ruiz}, {G{\'o}mez}, {Gu}, {Gurwell}, {Hada}, {Haggard},
  {Haworth}, {Hecht}, {Hesper}, {Heumann}, {Ho}, {Ho}, {Honma}, {Huang},
  {Huang}, {Hughes}, {Ikeda}, {Impellizzeri}, {Inoue}, {Issaoun}, {James},
  {Jannuzi}, {Janssen}, {Jeter}, {Jiang}, {Jim{\'e}nez-Rosales}, {Johnson},
  {Jorstad}, {Joshi}, {Jung}, {Karami}, {Karuppusamy}, {Kawashima}, {Keating},
  {Kettenis}, {Kim}, {Kim}, {Kim}, {Kim}, {Kino}, {Koay}, {Kocherlakota},
  {Kofuji}, {Koch}, {Koyama}, {Kramer}, {Kramer}, {Krichbaum}, {Kuo}, {La
  Bella}, {Lauer}, {Lee}, {Lee}, {Leung}, {Levis}, {Li}, {Lico}, {Lindahl},
  {Lindqvist}, {Lisakov}, {Liu}, {Liu}, {Liuzzo}, {Lo}, {Lobanov}, {Loinard},
  {Lonsdale}, {Lu}, {Mao}, {Marchili}, {Markoff}, {Marrone}, {Marscher},
  {Mart{\'\i}-Vidal}, {Matsushita}, {Matthews}, {Medeiros}, {Menten},
  {Michalik}, {Mizuno}, {Mizuno}, {Moran}, {Moriyama}, {Moscibrodzka},
  {M{\"u}ller}, {Mus}, {Musoke}, {Myserlis}, {Nadolski}, {Nagai}, {Nagar},
  {Nakamura}, {Narayan}, {Narayanan}, {Natarajan}, {Nathanail}, {Navarro
  Fuentes}, {Neilsen}, {Neri}, {Ni}, {Noutsos}, {Nowak}, {Oh}, {Okino},
  {Olivares}, {Ortiz-Le{\'o}n}, {Oyama}, {{\"O}zel}, {Palumbo}, {Paraschos},
  {Park}, {Parsons}, {Patel}, {Pen}, {Pesce}, {Pi{\'e}tu}, {Plambeck},
  {PopStefanija}, {Porth}, {P{\"o}tzl}, {Prather}, {Preciado-L{\'o}pez},
  {Psaltis}, {Pu}, {Ramakrishnan}, {Rao}, {Rawlings}, {Raymond}, {Rezzolla},
  {Ricarte}, {Ripperda}, {Roelofs}, {Rogers}, {Ros}, {Romero-Ca{\~n}izales},
  {Roshanineshat}, {Rottmann}, {Roy}, {Ruiz}, {Ruszczyk}, {Rygl},
  {S{\'a}nchez}, {S{\'a}nchez-Arg{\"u}elles}, {S{\'a}nchez-Portal}, {Sasada},
  {Satapathy}, {Savolainen}, {Schloerb}, {Schonfeld}, {Schuster}, {Shao},
  {Shen}, {Small}, {Sohn}, {SooHoo}, {Souccar}, {Sun}, {Tazaki}, {Tetarenko},
  {Tiede}, {Tilanus}, {Titus}, {Torne}, {Traianou}, {Trent}, {Trippe}, {Turk},
  {van Bemmel}, {van Langevelde}, {van Rossum}, {Vos}, {Wagner},
  {Ward-Thompson}, {Wardle}, {Weintroub}, {Wex}, {Wharton}, {Wielgus}, {Wiik},
  {Witzel}, {Wondrak}, {Wong}, {Wu}, {Yamaguchi}, {Yoon}, {Young}, {Young},
  {Younsi}, {Yuan}, {Yuan}, {Zensus}, {Zhang}, {Zhao}, {Zhao}, {Agurto},
  {Araneda}, {Arriagada}, {Bertarini}, {Berthold}, {Blanchard}, {Brown},
  {C{\'a}rdenas}, {Cantzler}, {Caro}, {Chuter}, {Ciechanowicz}, {Coulson},
  {Crowley}, {Degenaar}, {Dornbusch}, {Dur{\'a}n}, {Forster}, {Geertsema},
  {Gonz{\'a}lez}, {Graham}, {Gueth}, {Han}, {Herrera}, {Herrero-Illana},
  {Heyminck}, {Hoge}, {Huang}, {Jiang}, {John}, {Klein}, {Kubo}, {Kuroda},
  {Kwon}, {Laing}, {Liu}, {Liu}, {Mac-Auliffe}, {Martin-Cocher}, {Matulonis},
  {Messias}, {Meyer-Zhao}, {Montenegro-Montes}, {Montgomerie}, {Muders},
  {Nishioka}, {Norton}, {Olivares}, {P{\'e}rez-Beaupuits}, {Parra}, {Poirier},
  {Pradel}, {Raffin}, {Ram{\'\i}rez}, {Reynolds}, {Saez-Madain}, {Santana},
  {Silva}, {Sousa}, {Stahm}, {Torstensson}, {Venegas}, {Walther}, {Wieching},
  {Wijnands}, {Wouterloot}, \& {EHT Collaboration}}]{eht_sgra_II}
{Event Horizon Telescope Collaboration}, {Akiyama}, K., {Alberdi}, A., {et~al.}
  2022{\natexlab{b}}, \apjl, 930, L13

\bibitem[{{Event Horizon Telescope Collaboration}
  {et~al.}(2022{\natexlab{c}}){Event Horizon Telescope Collaboration},
  {Akiyama}, {Alberdi}, {Alef}, {Algaba}, {Anantua}, {Asada}, {Azulay}, {Bach},
  {Baczko}, {Ball}, {Balokovi{\'c}}, {Barrett}, {Baub{\"o}ck}, {Benson},
  {Bintley}, {Blackburn}, {Blundell}, {Bouman}, {Bower}, {Boyce}, {Bremer},
  {Brinkerink}, {Brissenden}, {Britzen}, {Broderick}, {Broguiere}, {Bronzwaer},
  {Bustamante}, {Byun}, {Carlstrom}, {Ceccobello}, {Chael}, {Chan},
  {Chatterjee}, {Chatterjee}, {Chen}, {Chen}, {Cheng}, {Cho}, {Christian},
  {Conroy}, {Conway}, {Cordes}, {Crawford}, {Crew}, {Cruz-Osorio}, {Cui},
  {Davelaar}, {De Laurentis}, {Deane}, {Dempsey}, {Desvignes}, {Dexter},
  {Dhruv}, {Doeleman}, {Dougal}, {Dzib}, {Eatough}, {Emami}, {Falcke}, {Farah},
  {Fish}, {Fomalont}, {Ford}, {Fraga-Encinas}, {Freeman}, {Friberg}, {Fromm},
  {Fuentes}, {Galison}, {Gammie}, {Garc{\'\i}a}, {Gentaz}, {Georgiev}, {Goddi},
  {Gold}, {G{\'o}mez-Ruiz}, {G{\'o}mez}, {Gu}, {Gurwell}, {Hada}, {Haggard},
  {Haworth}, {Hecht}, {Hesper}, {Heumann}, {Ho}, {Ho}, {Honma}, {Huang},
  {Huang}, {Hughes}, {Ikeda}, {Impellizzeri}, {Inoue}, {Issaoun}, {James},
  {Jannuzi}, {Janssen}, {Jeter}, {Jiang}, {Jim{\'e}nez-Rosales}, {Johnson},
  {Jorstad}, {Joshi}, {Jung}, {Karami}, {Karuppusamy}, {Kawashima}, {Keating},
  {Kettenis}, {Kim}, {Kim}, {Kim}, {Kim}, {Kino}, {Koay}, {Kocherlakota},
  {Kofuji}, {Koch}, {Koyama}, {Kramer}, {Kramer}, {Krichbaum}, {Kuo}, {La
  Bella}, {Lauer}, {Lee}, {Lee}, {Leung}, {Levis}, {Li}, {Lico}, {Lindahl},
  {Lindqvist}, {Lisakov}, {Liu}, {Liu}, {Liuzzo}, {Lo}, {Lobanov}, {Loinard},
  {Lonsdale}, {Lu}, {Mao}, {Marchili}, {Markoff}, {Marrone}, {Marscher},
  {Mart{\'\i}-Vidal}, {Matsushita}, {Matthews}, {Medeiros}, {Menten},
  {Michalik}, {Mizuno}, {Mizuno}, {Moran}, {Moriyama}, {Moscibrodzka},
  {M{\"u}ller}, {Mus}, {Musoke}, {Myserlis}, {Nadolski}, {Nagai}, {Nagar},
  {Nakamura}, {Narayan}, {Narayanan}, {Natarajan}, {Nathanail}, {Navarro
  Fuentes}, {Neilsen}, {Neri}, {Ni}, {Noutsos}, {Nowak}, {Oh}, {Okino},
  {Olivares}, {Ortiz-Le{\'o}n}, {Oyama}, {{\"O}zel}, {Palumbo}, {Paraschos},
  {Park}, {Parsons}, {Patel}, {Pen}, {Pesce}, {Pi{\'e}tu}, {Plambeck},
  {PopStefanija}, {Porth}, {P{\"o}tzl}, {Prather}, {Preciado-L{\'o}pez},
  {Psaltis}, {Pu}, {Ramakrishnan}, {Rao}, {Rawlings}, {Raymond}, {Rezzolla},
  {Ricarte}, {Ripperda}, {Roelofs}, {Rogers}, {Ros}, {Romero-Ca{\~n}izales},
  {Roshanineshat}, {Rottmann}, {Roy}, {Ruiz}, {Ruszczyk}, {Rygl},
  {S{\'a}nchez}, {S{\'a}nchez-Arg{\"u}elles}, {S{\'a}nchez-Portal}, {Sasada},
  {Satapathy}, {Savolainen}, {Schloerb}, {Schonfeld}, {Schuster}, {Shao},
  {Shen}, {Small}, {Sohn}, {SooHoo}, {Souccar}, {Sun}, {Tazaki}, {Tetarenko},
  {Tiede}, {Tilanus}, {Titus}, {Torne}, {Traianou}, {Trent}, {Trippe}, {Turk},
  {van Bemmel}, {van Langevelde}, {van Rossum}, {Vos}, {Wagner},
  {Ward-Thompson}, {Wardle}, {Weintroub}, {Wex}, {Wharton}, {Wielgus}, {Wiik},
  {Witzel}, {Wondrak}, {Wong}, {Wu}, {Yamaguchi}, {Yoon}, {Young}, {Young},
  {Younsi}, {Yuan}, {Yuan}, {Zensus}, {Zhang}, {Zhao}, {Zhao}, \& {EHT
  Collaboration}}]{eht_paper_III}
{Event Horizon Telescope Collaboration}, {Akiyama}, K., {Alberdi}, A., {et~al.}
  2022{\natexlab{c}}, \apjl, 930, L14

\bibitem[{{Event Horizon Telescope Collaboration}
  {et~al.}(2022{\natexlab{d}}){Event Horizon Telescope Collaboration},
  {Akiyama}, {Alberdi}, {Alef}, {Algaba}, {Anantua}, {Asada}, {Azulay}, {Bach},
  {Baczko}, {Ball}, {Balokovi{\'c}}, {Barrett}, {Baub{\"o}ck}, {Benson},
  {Bintley}, {Blackburn}, {Blundell}, {Bouman}, {Bower}, {Boyce}, {Bremer},
  {Brinkerink}, {Brissenden}, {Britzen}, {Broderick}, {Broguiere}, {Bronzwaer},
  {Bustamante}, {Byun}, {Carlstrom}, {Ceccobello}, {Chael}, {Chan},
  {Chatterjee}, {Chatterjee}, {Chen}, {Chen}, {Cheng}, {Cho}, {Christian},
  {Conroy}, {Conway}, {Cordes}, {Crawford}, {Crew}, {Cruz-Osorio}, {Cui},
  {Davelaar}, {De Laurentis}, {Deane}, {Dempsey}, {Desvignes}, {Dexter},
  {Dhruv}, {Doeleman}, {Dougal}, {Dzib}, {Eatough}, {Emami}, {Falcke}, {Farah},
  {Fish}, {Fomalont}, {Ford}, {Fraga-Encinas}, {Freeman}, {Friberg}, {Fromm},
  {Fuentes}, {Galison}, {Gammie}, {Garc{\'\i}a}, {Gentaz}, {Georgiev}, {Goddi},
  {Gold}, {G{\'o}mez-Ruiz}, {G{\'o}mez}, {Gu}, {Gurwell}, {Hada}, {Haggard},
  {Haworth}, {Hecht}, {Hesper}, {Heumann}, {Ho}, {Ho}, {Honma}, {Huang},
  {Huang}, {Hughes}, {Ikeda}, {Impellizzeri}, {Inoue}, {Issaoun}, {James},
  {Jannuzi}, {Janssen}, {Jeter}, {Jiang}, {Jim{\'e}nez-Rosales}, {Johnson},
  {Jorstad}, {Joshi}, {Jung}, {Karami}, {Karuppusamy}, {Kawashima}, {Keating},
  {Kettenis}, {Kim}, {Kim}, {Kim}, {Kim}, {Kino}, {Koay}, {Kocherlakota},
  {Kofuji}, {Koch}, {Koyama}, {Kramer}, {Kramer}, {Krichbaum}, {Kuo}, {La
  Bella}, {Lauer}, {Lee}, {Lee}, {Leung}, {Levis}, {Li}, {Lico}, {Lindahl},
  {Lindqvist}, {Lisakov}, {Liu}, {Liu}, {Liuzzo}, {Lo}, {Lobanov}, {Loinard},
  {Lonsdale}, {Lu}, {Mao}, {Marchili}, {Markoff}, {Marrone}, {Marscher},
  {Mart{\'\i}-Vidal}, {Matsushita}, {Matthews}, {Medeiros}, {Menten},
  {Michalik}, {Mizuno}, {Mizuno}, {Moran}, {Moriyama}, {Moscibrodzka},
  {M{\"u}ller}, {Mus}, {Musoke}, {Myserlis}, {Nadolski}, {Nagai}, {Nagar},
  {Nakamura}, {Narayan}, {Narayanan}, {Natarajan}, {Nathanail}, {Navarro
  Fuentes}, {Neilsen}, {Neri}, {Ni}, {Noutsos}, {Nowak}, {Oh}, {Okino},
  {Olivares}, {Ortiz-Le{\'o}n}, {Oyama}, {Palumbo}, {Paraschos}, {Park},
  {Parsons}, {Patel}, {Pen}, {Pesce}, {Pi{\'e}tu}, {Plambeck}, {PopStefanija},
  {Porth}, {P{\"o}tzl}, {Prather}, {Preciado-L{\'o}pez}, {Pu}, {Ramakrishnan},
  {Rao}, {Rawlings}, {Raymond}, {Rezzolla}, {Ricarte}, {Ripperda}, {Roelofs},
  {Rogers}, {Ros}, {Romero-Ca{\~n}izales}, {Roshanineshat}, {Rottmann}, {Roy},
  {Ruiz}, {Ruszczyk}, {Rygl}, {S{\'a}nchez}, {S{\'a}nchez-Arg{\"u}elles},
  {S{\'a}nchez-Portal}, {Sasada}, {Satapathy}, {Savolainen}, {Schloerb},
  {Schonfeld}, {Schuster}, {Shao}, {Shen}, {Small}, {Sohn}, {SooHoo},
  {Souccar}, {Sun}, {Tazaki}, {Tetarenko}, {Tiede}, {Tilanus}, {Titus},
  {Torne}, {Traianou}, {Trent}, {Trippe}, {Turk}, {van Bemmel}, {van
  Langevelde}, {van Rossum}, {Vos}, {Wagner}, {Ward-Thompson}, {Wardle},
  {Weintroub}, {Wex}, {Wharton}, {Wielgus}, {Wiik}, {Witzel}, {Wondrak},
  {Wong}, {Wu}, {Yamaguchi}, {Yoon}, {Young}, {Young}, {Younsi}, {Yuan},
  {Yuan}, {Zensus}, {Zhang}, {Zhao}, {Zhao}, {Chang}, \& {EHT
  Collaboration}}]{eht_paper_IV}
{Event Horizon Telescope Collaboration}, {Akiyama}, K., {Alberdi}, A., {et~al.}
  2022{\natexlab{d}}, \apjl, 930, L15

\bibitem[{{Event Horizon Telescope Collaboration}
  {et~al.}(2022{\natexlab{e}}){Event Horizon Telescope Collaboration},
  {Akiyama}, {Alberdi}, {Alef}, {Algaba}, {Anantua}, {Asada}, {Azulay}, {Bach},
  {Baczko}, {Ball}, {Balokovi{\'c}}, {Barrett}, {Baub{\"o}ck}, {Benson},
  {Bintley}, {Blackburn}, {Blundell}, {Bouman}, {Bower}, {Boyce}, {Bremer},
  {Brinkerink}, {Brissenden}, {Britzen}, {Broderick}, {Broguiere}, {Bronzwaer},
  {Bustamante}, {Byun}, {Carlstrom}, {Ceccobello}, {Chael}, {Chan},
  {Chatterjee}, {Chatterjee}, {Chen}, {Chen}, {Cheng}, {Cho}, {Christian},
  {Conroy}, {Conway}, {Cordes}, {Crawford}, {Crew}, {Cruz-Osorio}, {Cui},
  {Davelaar}, {De Laurentis}, {Deane}, {Dempsey}, {Desvignes}, {Dexter},
  {Dhruv}, {Doeleman}, {Dougal}, {Dzib}, {Eatough}, {Emami}, {Falcke}, {Farah},
  {Fish}, {Fomalont}, {Ford}, {Fraga-Encinas}, {Freeman}, {Friberg}, {Fromm},
  {Fuentes}, {Galison}, {Gammie}, {Garc{\'\i}a}, {Gentaz}, {Georgiev}, {Goddi},
  {Gold}, {G{\'o}mez-Ruiz}, {G{\'o}mez}, {Gu}, {Gurwell}, {Hada}, {Haggard},
  {Haworth}, {Hecht}, {Hesper}, {Heumann}, {Ho}, {Ho}, {Honma}, {Huang},
  {Huang}, {Hughes}, {Ikeda}, {Impellizzeri}, {Inoue}, {Issaoun}, {James},
  {Jannuzi}, {Janssen}, {Jeter}, {Jiang}, {Jim{\'e}nez-Rosales}, {Johnson},
  {Jorstad}, {Joshi}, {Jung}, {Karami}, {Karuppusamy}, {Kawashima}, {Keating},
  {Kettenis}, {Kim}, {Kim}, {Kim}, {Kim}, {Kino}, {Koay}, {Kocherlakota},
  {Kofuji}, {Koch}, {Koyama}, {Kramer}, {Kramer}, {Krichbaum}, {Kuo}, {La
  Bella}, {Lauer}, {Lee}, {Lee}, {Leung}, {Levis}, {Li}, {Lico}, {Lindahl},
  {Lindqvist}, {Lisakov}, {Liu}, {Liu}, {Liuzzo}, {Lo}, {Lobanov}, {Loinard},
  {Lonsdale}, {Lu}, {Mao}, {Marchili}, {Markoff}, {Marrone}, {Marscher},
  {Mart{\'\i}-Vidal}, {Matsushita}, {Matthews}, {Medeiros}, {Menten},
  {Michalik}, {Mizuno}, {Mizuno}, {Moran}, {Moriyama}, {Moscibrodzka},
  {M{\"u}ller}, {Mus}, {Musoke}, {Myserlis}, {Nadolski}, {Nagai}, {Nagar},
  {Nakamura}, {Narayan}, {Narayanan}, {Natarajan}, {Nathanail}, {Navarro
  Fuentes}, {Neilsen}, {Neri}, {Ni}, {Noutsos}, {Nowak}, {Oh}, {Okino},
  {Olivares}, {Ortiz-Le{\'o}n}, {Oyama}, {{\"O}zel}, {Palumbo}, {Paraschos},
  {Park}, {Parsons}, {Patel}, {Pen}, {Pesce}, {Pi{\'e}tu}, {Plambeck},
  {PopStefanija}, {Porth}, {P{\"o}tzl}, {Prather}, {Preciado-L{\'o}pez},
  {Psaltis}, {Pu}, {Ramakrishnan}, {Rao}, {Rawlings}, {Raymond}, {Rezzolla},
  {Ricarte}, {Ripperda}, {Roelofs}, {Rogers}, {Ros}, {Romero-Ca{\~n}izales},
  {Roshanineshat}, {Rottmann}, {Roy}, {Ruiz}, {Ruszczyk}, {Rygl},
  {S{\'a}nchez}, {S{\'a}nchez-Arg{\"u}elles}, {S{\'a}nchez-Portal}, {Sasada},
  {Satapathy}, {Savolainen}, {Schloerb}, {Schonfeld}, {Schuster}, {Shao},
  {Shen}, {Small}, {Sohn}, {SooHoo}, {Souccar}, {Sun}, {Tazaki}, {Tetarenko},
  {Tiede}, {Tilanus}, {Titus}, {Torne}, {Traianou}, {Trent}, {Trippe}, {Turk},
  {van Bemmel}, {van Langevelde}, {van Rossum}, {Vos}, {Wagner},
  {Ward-Thompson}, {Wardle}, {Weintroub}, {Wex}, {Wharton}, {Wielgus}, {Wiik},
  {Witzel}, {Wondrak}, {Wong}, {Wu}, {Yamaguchi}, {Yoon}, {Young}, {Young},
  {Younsi}, {Yuan}, {Yuan}, {Zensus}, {Zhang}, {Zhao}, {Zhao}, {Chan}, {Qiu},
  {Ressler}, {White}, \& {EHT Collaboration}}]{eht_paper_V}
{Event Horizon Telescope Collaboration}, {Akiyama}, K., {Alberdi}, A., {et~al.}
  2022{\natexlab{e}}, \apjl, 930, L16

\bibitem[{{Event Horizon Telescope Collaboration}
  {et~al.}(2022{\natexlab{f}}){Event Horizon Telescope Collaboration},
  {Akiyama}, {Alberdi}, {Alef}, {Algaba}, {Anantua}, {Asada}, {Azulay}, {Bach},
  {Baczko}, {Ball}, {Balokovi{\'c}}, {Barrett}, {Baub{\"o}ck}, {Benson},
  {Bintley}, {Blackburn}, {Blundell}, {Bouman}, {Bower}, {Boyce}, {Bremer},
  {Brinkerink}, {Brissenden}, {Britzen}, {Broderick}, {Broguiere}, {Bronzwaer},
  {Bustamante}, {Byun}, {Carlstrom}, {Ceccobello}, {Chael}, {Chan},
  {Chatterjee}, {Chatterjee}, {Chen}, {Chen}, {Cheng}, {Cho}, {Christian},
  {Conroy}, {Conway}, {Cordes}, {Crawford}, {Crew}, {Cruz-Osorio}, {Cui},
  {Davelaar}, {De Laurentis}, {Deane}, {Dempsey}, {Desvignes}, {Dexter},
  {Dhruv}, {Doeleman}, {Dougal}, {Dzib}, {Eatough}, {Emami}, {Falcke}, {Farah},
  {Fish}, {Fomalont}, {Ford}, {Fraga-Encinas}, {Freeman}, {Friberg}, {Fromm},
  {Fuentes}, {Galison}, {Gammie}, {Garc{\'\i}a}, {Gentaz}, {Georgiev}, {Goddi},
  {Gold}, {G{\'o}mez-Ruiz}, {G{\'o}mez}, {Gu}, {Gurwell}, {Hada}, {Haggard},
  {Haworth}, {Hecht}, {Hesper}, {Heumann}, {Ho}, {Ho}, {Honma}, {Huang},
  {Huang}, {Hughes}, {Ikeda}, {Impellizzeri}, {Inoue}, {Issaoun}, {James},
  {Jannuzi}, {Janssen}, {Jeter}, {Jiang}, {Jim{\'e}nez-Rosales}, {Johnson},
  {Jorstad}, {Joshi}, {Jung}, {Karami}, {Karuppusamy}, {Kawashima}, {Keating},
  {Kettenis}, {Kim}, {Kim}, {Kim}, {Kim}, {Kino}, {Koay}, {Kocherlakota},
  {Kofuji}, {Koch}, {Koyama}, {Kramer}, {Kramer}, {Krichbaum}, {Kuo}, {Bella},
  {Lauer}, {Lee}, {Lee}, {Leung}, {Levis}, {Li}, {Lico}, {Lindahl},
  {Lindqvist}, {Lisakov}, {Liu}, {Liu}, {Liuzzo}, {Lo}, {Lobanov}, {Loinard},
  {Lonsdale}, {Lu}, {Mao}, {Marchili}, {Markoff}, {Marrone}, {Marscher},
  {Mart{\'\i}-Vidal}, {Matsushita}, {Matthews}, {Medeiros}, {Menten},
  {Michalik}, {Mizuno}, {Mizuno}, {Moran}, {Moriyama}, {Moscibrodzka},
  {M{\"u}ller}, {Mus}, {Musoke}, {Myserlis}, {Nadolski}, {Nagai}, {Nagar},
  {Nakamura}, {Narayan}, {Narayanan}, {Natarajan}, {Nathanail}, {Fuentes},
  {Neilsen}, {Neri}, {Ni}, {Noutsos}, {Nowak}, {Oh}, {Okino}, {Olivares},
  {Ortiz-Le{\'o}n}, {Oyama}, {{\"O}zel}, {Palumbo}, {Paraschos}, {Park},
  {Parsons}, {Patel}, {Pen}, {Pesce}, {Pi{\'e}tu}, {Plambeck}, {PopStefanija},
  {Porth}, {P{\"o}tzl}, {Prather}, {Preciado-L{\'o}pez}, {Psaltis}, {Pu},
  {Ramakrishnan}, {Rao}, {Rawlings}, {Raymond}, {Rezzolla}, {Ricarte},
  {Ripperda}, {Roelofs}, {Rogers}, {Ros}, {Romero-Ca{\~n}izales},
  {Roshanineshat}, {Rottmann}, {Roy}, {Ruiz}, {Ruszczyk}, {Rygl},
  {S{\'a}nchez}, {S{\'a}nchez-Arg{\"u}elles}, {S{\'a}nchez-Portal}, {Sasada},
  {Satapathy}, {Savolainen}, {Schloerb}, {Schonfeld}, {Schuster}, {Shao},
  {Shen}, {Small}, {Sohn}, {SooHoo}, {Souccar}, {Sun}, {Tazaki}, {Tetarenko},
  {Tiede}, {Tilanus}, {Titus}, {Torne}, {Traianou}, {Trent}, {Trippe}, {Turk},
  {van Bemmel}, {van Langevelde}, {van Rossum}, {Vos}, {Wagner},
  {Ward-Thompson}, {Wardle}, {Weintroub}, {Wex}, {Wharton}, {Wielgus}, {Wiik},
  {Witzel}, {Wondrak}, {Wong}, {Wu}, {Yamaguchi}, {Yoon}, {Young}, {Young},
  {Younsi}, {Yuan}, {Yuan}, {Zensus}, {Zhang}, {Zhao}, \&
  {Zhao}}]{eht_paper_VI}
{Event Horizon Telescope Collaboration}, {Akiyama}, K., {Alberdi}, A., {et~al.}
  2022{\natexlab{f}}, \apjl, 930, L17

\bibitem[{{Falcke} \& {Markoff}(2000)}]{Falcke2000}
{Falcke}, H. \& {Markoff}, S. 2000, \aap, 362, 113

\bibitem[{{Falcke} {et~al.}(2000){Falcke}, {Melia}, \&
  {Agol}}]{Falcke2000_bhshadow}
{Falcke}, H., {Melia}, F., \& {Agol}, E. 2000, \apjl, 528, L13

\bibitem[{{Fritz} {et~al.}(2011){Fritz}, {Gillessen}, {Dodds-Eden}, {Lutz},
  {Genzel}, {Raab}, {Ott}, {Pfuhl}, {Eisenhauer}, \& {Yusef-Zadeh}}]{Fritz2011}
{Fritz}, T.~K., {Gillessen}, S., {Dodds-Eden}, K., {et~al.} 2011, \apj, 737, 73

\bibitem[{{Galishnikova} {et~al.}(2022){Galishnikova}, {Philippov}, {Quataert},
  {Bacchini}, {Parfrey}, \& {Ripperda}}]{Galishnikova2022}
{Galishnikova}, A., {Philippov}, A., {Quataert}, E., {et~al.} 2022, arXiv
  e-prints, arXiv:2212.02583

\bibitem[{{Genzel} {et~al.}(2010){Genzel}, {Eisenhauer}, \&
  {Gillessen}}]{Genzel2010}
{Genzel}, R., {Eisenhauer}, F., \& {Gillessen}, S. 2010, Reviews of Modern
  Physics, 82, 3121

\bibitem[{{Genzel} {et~al.}(2003){Genzel}, {Sch{\"o}del}, {Ott}, {Eckart},
  {Alexander}, {Lacombe}, {Rouan}, \& {Aschenbach}}]{Genzel2003}
{Genzel}, R., {Sch{\"o}del}, R., {Ott}, T., {et~al.} 2003, \nat, 425, 934

\bibitem[{Gillessen {et~al.}(2006)Gillessen, Eisenhauer, Quataert, Genzel,
  Paumard, Trippe, Ott, Abuter, Eckart, Lagage, Lehnert, Tacconi, \&
  Martins}]{Gillessen2006}
Gillessen, S., Eisenhauer, F., Quataert, E., {et~al.} 2006, \apj, 640, 163

\bibitem[{{GRAVITY Collaboration} {et~al.}(2021){GRAVITY Collaboration},
  {Abuter}, {Amorim}, {Baub{\"o}ck}, {Baganoff}, {Berger}, {Boyce}, {Bonnet},
  {Brandner}, {Cl{\'e}net}, {Davies}, {de Zeeuw}, {Dexter}, {Dallilar},
  {Drescher}, {Eckart}, {Eisenhauer}, {Fazio}, {F{\"o}rster Schreiber},
  {Foster}, {Gammie}, {Garcia}, {Gao}, {Gendron}, {Genzel}, {Ghisellini},
  {Gillessen}, {Gurwell}, {Habibi}, {Haggard}, {Hailey}, {Harrison}, {Haubois},
  {Hei{\ss}el}, {Henning}, {Hippler}, {Hora}, {Horrobin},
  {Jim{\'e}nez-Rosales}, {Jochum}, {Jocou}, {Kaufer}, {Kervella}, {Lacour},
  {Lapeyr{\`e}re}, {Le Bouquin}, {L{\'e}na}, {Lowrance}, {Lutz}, {Markoff},
  {Mori}, {Morris}, {Neilsen}, {Nowak}, {Ott}, {Paumard}, {Perraut}, {Perrin},
  {Ponti}, {Pfuhl}, {Rabien}, {Rodr{\'\i}guez-Coira}, {Shangguan}, {Shimizu},
  {Scheithauer}, {Smith}, {Stadler}, {Stern}, {Straub}, {Straubmeier}, {Sturm},
  {Tacconi}, {Vincent}, {von Fellenberg}, {Waisberg}, {Widmann}, {Wieprecht},
  {Wiezorrek}, {Willner}, {Witzel}, {Woillez}, {Yazici}, {Young}, {Zhang}, \&
  {Zins}}]{GravityCollaboration2021_xrayflare}
{GRAVITY Collaboration}, {Abuter}, R., {Amorim}, A., {et~al.} 2021, \aap, 654,
  A22

\bibitem[{{GRAVITY Collaboration} {et~al.}(2020{\natexlab{a}}){GRAVITY
  Collaboration}, Abuter, Amorim, Baub{\"{o}}ck, Berger, Bonnet, Brandner,
  Cardoso, Cl{\'{e}}net, {De Zeeuw}, Dallilar, Dexter, Eckart, Eisenhauer,
  {F{\"{o}}rster Schreiber}, Garcia, Gao, Gendron, Genzel, Gillessen, Habibi,
  Haubois, Henning, Hippler, Horrobin, Jim{\'{e}}nez-Rosales, Jochum, Jocou,
  Kaufer, Kervella, Lacour, Lapeyr{\`{e}}re, {Le Bouquin}, L{\'{e}}na, Nowak,
  Ott, Paumard, Perraut, Perrin, Pfuhl, Ponti, {Rodriguez Coira}, Shangguan,
  Scheithauer, Stadler, Straub, Straubmeier, Sturm, Tacconi, Vincent, {Von
  Fellenberg}, Waisberg, Widmann, Wieprecht, Wiezorrek, Woillez, Yazici, \&
  Zins}]{GRAVITYCollaboration2020flux}
{GRAVITY Collaboration}, Abuter, R., Amorim, A., {et~al.} 2020{\natexlab{a}},
  Astronomy and Astrophysics, 638

\bibitem[{{GRAVITY Collaboration} {et~al.}(2018){GRAVITY Collaboration},
  Abuter, Amorim, Baub{\"{o}}ck, Berger, Bonnet, Brandner, Cl{\'{e}}net,
  {Coud{\'{e}} du Foresto}, de~Zeeuw, Deen, Dexter, Duvert, Eckart, Eisenhauer,
  {F{\"{o}}rster Schreiber}, Garcia, Gao, Gendron, Genzel, Gillessen, Guajardo,
  Habibi, Haubois, Henning, Hippler, Horrobin, Huber, Jim{\'{e}}nez-Rosales,
  Jocou, Kervella, Lacour, Lapeyr{\`{e}}re, Lazareff, {Le Bouquin}, L{\'{e}}na,
  Lippa, Ott, Panduro, Paumard, Perraut, Perrin, Pfuhl, Plewa, Rabien,
  Rodr{\'{i}}guez-Coira, Rousset, Sternberg, Straub, Straubmeier, Sturm,
  Tacconi, Vincent, von Fellenberg, Waisberg, Widmann, Wieprecht, Wiezorrek,
  Woillez, \& Yazici}]{GRAVITYCollaboration2018_orbital}
{GRAVITY Collaboration}, Abuter, R., Amorim, A., {et~al.} 2018, Astronomy {\&}
  Astrophysics, 618, L10

\bibitem[{{GRAVITY Collaboration} {et~al.}(2020{\natexlab{b}}){GRAVITY
  Collaboration}, {Baub{\"o}ck}, {Dexter}, {Abuter}, {Amorim}, {Berger},
  {Bonnet}, {Brandner}, {Cl{\'e}net}, {Coud{\'e} Du Foresto}, {de Zeeuw},
  {Duvert}, {Eckart}, {Eisenhauer}, {F{\"o}rster Schreiber}, {Gao}, {Garcia},
  {Gendron}, {Genzel}, {Gerhard}, {Gillessen}, {Habibi}, {Haubois}, {Henning},
  {Hippler}, {Horrobin}, {Jim{\'e}nez-Rosales}, {Jocou}, {Kervella}, {Lacour},
  {Lapeyr{\`e}re}, {Le Bouquin}, {L{\'e}na}, {Ott}, {Paumard}, {Perraut},
  {Perrin}, {Pfuhl}, {Rabien}, {Rodriguez Coira}, {Rousset}, {Scheithauer},
  {Stadler}, {Sternberg}, {Straub}, {Straubmeier}, {Sturm}, {Tacconi},
  {Vincent}, {von Fellenberg}, {Waisberg}, {Widmann}, {Wieprecht}, {Wiezorrek},
  {Woillez}, \& {Yazici}}]{GravityCollaboration2020_orbital}
{GRAVITY Collaboration}, {Baub{\"o}ck}, M., {Dexter}, J., {et~al.}
  2020{\natexlab{b}}, \aap, 635, A143

\bibitem[{{GRAVITY Collaboration} {et~al.}(2020{\natexlab{c}}){GRAVITY
  Collaboration}, {Jim{\'e}nez-Rosales}, {Dexter}, {Widmann}, {Baub{\"o}ck},
  {Abuter}, {Amorim}, {Berger}, {Bonnet}, {Brandner}, {Cl{\'e}net}, {de Zeeuw},
  {Eckart}, {Eisenhauer}, {F{\"o}rster Schreiber}, {Garcia}, {Gao}, {Gendron},
  {Genzel}, {Gillessen}, {Habibi}, {Haubois}, {Hei{\ss}el}, {Henning},
  {Hippler}, {Horrobin}, {Jochum}, {Jocou}, {Kaufer}, {Kervella}, {Lacour},
  {Lapeyr{\`e}re}, {Le Bouquin}, {L{\'e}na}, {Nowak}, {Ott}, {Paumard},
  {Perraut}, {Perrin}, {Pfuhl}, {Rodr{\'\i}guez-Coira}, {Shangguan},
  {Scheithauer}, {Stadler}, {Straub}, {Straubmeier}, {Sturm}, {Tacconi},
  {Vincent}, {von Fellenberg}, {Waisberg}, {Wieprecht}, {Wiezorrek}, {Woillez},
  {Yazici}, \& {Zins}}]{GravityCollaboration2020_polariflares}
{GRAVITY Collaboration}, {Jim{\'e}nez-Rosales}, A., {Dexter}, J., {et~al.}
  2020{\natexlab{c}}, \aap, 643, A56

\bibitem[{{GRAVITY Collaboration} {et~al.}(2020{\natexlab{d}}){GRAVITY
  Collaboration}, {Jim{\'e}nez-Rosales}, {Dexter}, {Widmann}, {Baub{\"o}ck},
  {Abuter}, {Amorim}, {Berger}, {Bonnet}, {Brandner}, {Cl{\'e}net}, {de Zeeuw},
  {Eckart}, {Eisenhauer}, {F{\"o}rster Schreiber}, {Garcia}, {Gao}, {Gendron},
  {Genzel}, {Gillessen}, {Habibi}, {Haubois}, {Hei{\ss}el}, {Henning},
  {Hippler}, {Horrobin}, {Jochum}, {Jocou}, {Kaufer}, {Kervella}, {Lacour},
  {Lapeyr{\`e}re}, {Le Bouquin}, {L{\'e}na}, {Nowak}, {Ott}, {Paumard},
  {Perraut}, {Perrin}, {Pfuhl}, {Rodr{\'\i}guez-Coira}, {Shangguan},
  {Scheithauer}, {Stadler}, {Straub}, {Straubmeier}, {Sturm}, {Tacconi},
  {Vincent}, {von Fellenberg}, {Waisberg}, {Wieprecht}, {Wiezorrek}, {Woillez},
  {Yazici}, \& {Zins}}]{GravityCollaboration2020_polarization_ale}
{GRAVITY Collaboration}, {Jim{\'e}nez-Rosales}, A., {Dexter}, J., {et~al.}
  2020{\natexlab{d}}, \aap, 643, A56

\bibitem[{{Hamaus} {et~al.}(2009){Hamaus}, {Paumard}, {M{\"u}ller},
  {Gillessen}, {Eisenhauer}, {Trippe}, \& {Genzel}}]{Hamaus2009}
{Hamaus}, N., {Paumard}, T., {M{\"u}ller}, T., {et~al.} 2009, \apj, 692, 902

\bibitem[{Hora {et~al.}(2014)Hora, Witzel, Ashby, Becklin, Carey, Fazio, Ghez,
  Ingalls, Meyer, Morris, Smith, \& Willner}]{Hora2014}
Hora, J.~L., Witzel, G., Ashby, M.~L., {et~al.} 2014, Astrophysical Journal,
  793, 120

\bibitem[{{Hornstein} {et~al.}(2007){Hornstein}, {Matthews}, {Ghez}, {Lu},
  {Morris}, {Becklin}, {Rafelski}, \& {Baganoff}}]{Hornstein2007}
{Hornstein}, S.~D., {Matthews}, K., {Ghez}, A.~M., {et~al.} 2007, \apj, 667,
  900

\bibitem[{{Karssen} {et~al.}(2017){Karssen}, {Bursa}, {Eckart}, {Valencia-S},
  {Dov{\v{c}}iak}, {Karas}, \& {Hor{\'a}k}}]{Karsen2017}
{Karssen}, G.~D., {Bursa}, M., {Eckart}, A., {et~al.} 2017, \mnras, 472, 4422

\bibitem[{{Krabbe} {et~al.}(2006){Krabbe}, {Iserlohe}, {Larkin}, {Barczys},
  {McElwain}, {Weiss}, {Wright}, \& {Quirrenbach}}]{Krabbe2006}
{Krabbe}, A., {Iserlohe}, C., {Larkin}, J.~E., {et~al.} 2006, \apjl, 642, L145

\bibitem[{{Liu} {et~al.}(2016){Liu}, {Wright}, {Zhao}, {Mills},
  {Requena-Torres}, {Matsushita}, {Mart{\'{\i}}n}, {Ott}, {Morris}, {Longmore},
  {Brinkerink}, \& {Falcke}}]{Liu2016}
{Liu}, H.~B., {Wright}, M.~C.~H., {Zhao}, J.-H., {et~al.} 2016, \aap, 593, A44

\bibitem[{{Lu} {et~al.}(2018){Lu}, {Krichbaum}, {Roy}, {Fish}, {Doeleman},
  {Johnson}, {Akiyama}, {Psaltis}, {Alef}, {Asada}, {Beaudoin}, {Bertarini},
  {Blackburn}, {Blundell}, {Bower}, {Brinkerink}, {Broderick}, {Cappallo},
  {Crew}, {Dexter}, {Dexter}, {Falcke}, {Freund}, {Friberg}, {Greer},
  {Gurwell}, {Ho}, {Honma}, {Inoue}, {Kim}, {Lamb}, {Lindqvist}, {Macmahon},
  {Marrone}, {Mart{\'\i}-Vidal}, {Menten}, {Moran}, {Nagar}, {Plambeck},
  {Primiani}, {Rogers}, {Ros}, {Rottmann}, {SooHoo}, {Spilker}, {Stone},
  {Strittmatter}, {Tilanus}, {Titus}, {Vertatschitsch}, {Wagner}, {Weintroub},
  {Wright}, {Young}, {Zensus}, \& {Ziurys}}]{Lu2018}
{Lu}, R.-S., {Krichbaum}, T.~P., {Roy}, A.~L., {et~al.} 2018, \apj, 859, 60

\bibitem[{{Mao} {et~al.}(2017){Mao}, {Dexter}, \& {Quataert}}]{Mao2017}
{Mao}, S.~A., {Dexter}, J., \& {Quataert}, E. 2017, \mnras, 466, 4307

\bibitem[{{Marrone} {et~al.}(2006){Marrone}, {Moran}, {Zhao}, \&
  {Rao}}]{Marrone2006}
{Marrone}, D.~P., {Moran}, J.~M., {Zhao}, J.-H., \& {Rao}, R. 2006, in Journal
  of Physics Conference Series, Vol.~54, Journal of Physics Conference Series,
  ed. R.~{Sch{\"o}del}, G.~C. {Bower}, M.~P. {Muno}, S.~{Nayakshin}, \&
  T.~{Ott}, 354--362

\bibitem[{{Meyer} {et~al.}(2014){Meyer}, {Witzel}, {Longstaff}, \&
  {Ghez}}]{Meyer2014}
{Meyer}, L., {Witzel}, G., {Longstaff}, F.~A., \& {Ghez}, A.~M. 2014, \apj,
  791, 24

\bibitem[{{Michail} {et~al.}(2021){Michail}, {Wardle}, {Yusef-Zadeh}, \&
  {Kunneriath}}]{Wardle2021}
{Michail}, J.~M., {Wardle}, M., {Yusef-Zadeh}, F., \& {Kunneriath}, D. 2021,
  \apj, 923, 54

\bibitem[{{Morris} {et~al.}(2012){Morris}, {Meyer}, \& {Ghez}}]{Morris2012}
{Morris}, M.~R., {Meyer}, L., \& {Ghez}, A.~M. 2012, Research in Astronomy and
  Astrophysics, 12, 995

\bibitem[{{Mo{\'s}cibrodzka} \& {Falcke}(2013)}]{Moscibrodzka2013}
{Mo{\'s}cibrodzka}, M. \& {Falcke}, H. 2013, \aap, 559, L3

\bibitem[{{N{\"a}ttil{\"a}} \& {Beloborodov}(2021)}]{Nattila2021}
{N{\"a}ttil{\"a}}, J. \& {Beloborodov}, A.~M. 2021, \apj, 921, 87

\bibitem[{{Pearson}(1901)}]{Pearson1901}
{Pearson}, K. 1901, The London, Edinburgh, and Dublin Philosophical Magazine
  and Journal of Science, 2, 559–572

\bibitem[{{Ponti} {et~al.}(2017){Ponti}, {George}, {Scaringi}, {Zhang}, {Jin},
  {Dexter}, {Terrier}, {Clavel}, {Degenaar}, {Eisenhauer}, {Genzel},
  {Gillessen}, {Goldwurm}, {Habibi}, {Haggard}, {Hailey}, {Harrison},
  {Merloni}, {Mori}, {Nandra}, {Ott}, {Pfuhl}, {Plewa}, \&
  {Waisberg}}]{Ponti2017}
{Ponti}, G., {George}, E., {Scaringi}, S., {et~al.} 2017, \mnras, 468, 2447

\bibitem[{{Porth} {et~al.}(2021){Porth}, {Mizuno}, {Younsi}, \&
  {Fromm}}]{Porth2021}
{Porth}, O., {Mizuno}, Y., {Younsi}, Z., \& {Fromm}, C.~M. 2021, \mnras, 502,
  2023

\bibitem[{Press {et~al.}(1992)Press, Teukolsky, Vetterling, \&
  Flannery}]{PresTeukVettFlan92}
Press, W.~H., Teukolsky, S.~A., Vetterling, W.~T., \& Flannery, B.~P. 1992,
  Numerical Recipes in C, 2nd edn. (Cambridge, USA: Cambridge University Press)

\bibitem[{{Priestley}(1988)}]{Priestley1988}
{Priestley}, M.~B. 1988, {Non-linear and non-stationary time series analysis}
  (Academic Press), 237

\bibitem[{{Ressler} {et~al.}(2020){Ressler}, {White}, {Quataert}, \&
  {Stone}}]{Ressler2020}
{Ressler}, S.~M., {White}, C.~J., {Quataert}, E., \& {Stone}, J.~M. 2020,
  \apjl, 896, L6

\bibitem[{{Ripperda} {et~al.}(2020){Ripperda}, {Bacchini}, \&
  {Philippov}}]{Ripperda2020}
{Ripperda}, B., {Bacchini}, F., \& {Philippov}, A.~A. 2020, \apj, 900, 100

\bibitem[{{Ripperda} {et~al.}(2022){Ripperda}, {Liska}, {Chatterjee}, {Musoke},
  {Philippov}, {Markoff}, {Tchekhovskoy}, \& {Younsi}}]{Ripperda2022}
{Ripperda}, B., {Liska}, M., {Chatterjee}, K., {et~al.} 2022, \apjl, 924, L32

\bibitem[{{Risken}(1989)}]{Risken1989}
{Risken}, H. 1989, {The Fokker-Planck equation. Methods of solution and
  applications} (Springer), 427

\bibitem[{{Scargle}(1981)}]{Scargle1981}
{Scargle}, J.~D. 1981, \apjs, 45, 1

\bibitem[{{Scargle}(2020)}]{Scargle2020}
{Scargle}, J.~D. 2020, \apj, 895, 90

\bibitem[{{Skilling}(2004)}]{Skilling2004_dynesty}
{Skilling}, J. 2004, in American Institute of Physics Conference Series, Vol.
  735, Bayesian Inference and Maximum Entropy Methods in Science and
  Engineering: 24th International Workshop on Bayesian Inference and Maximum
  Entropy Methods in Science and Engineering, ed. R.~{Fischer}, R.~{Preuss}, \&
  U.~V. {Toussaint}, 395--405

\bibitem[{Skilling(2006{\natexlab{a}})}]{Skilling2006_dynesty}
Skilling, J. 2006{\natexlab{a}}, Bayesian Analysis, 1, 833

\bibitem[{Skilling(2006{\natexlab{b}})}]{Skilling2006_dynesty2}
Skilling, J. 2006{\natexlab{b}}, Bayesian Analysis, 1, 833

\bibitem[{{Speagle}(2020)}]{Speagle2020_dynesty}
{Speagle}, J.~S. 2020, \mnras, 493, 3132

\bibitem[{{Uhlenbeck} \& {Ornstein}(1930)}]{Uhlenbeck1930}
{Uhlenbeck}, G.~E. \& {Ornstein}, L.~S. 1930, Physical Review, 36, 823

\bibitem[{von Fellenberg {et~al.}(2018)von Fellenberg, Gillessen,
  Graci{\'{a}}-Carpio, Fritz, Dexter, Baub{\"{o}}ck, Ponti, Gao, Habibi, Plewa,
  Pfuhl, Jimenez-Rosales, Waisberg, Widmann, Ott, Eisenhauer, \&
  Genzel}]{VonFellenberg2018}
von Fellenberg, S.~D., Gillessen, S., Graci{\'{a}}-Carpio, J., {et~al.} 2018,
  \apj, 862, 129

\bibitem[{{Werner} \& {Uzdensky}(2021)}]{Werner2021}
{Werner}, G.~R. \& {Uzdensky}, D.~A. 2021, Journal of Plasma Physics, 87,
  905870613

\bibitem[{{Wielgus} {et~al.}(2022){Wielgus}, {Moscibrodzka}, {Vos}, {Gelles},
  {Mart{\'\i}-Vidal}, {Farah}, {Marchili}, {Goddi}, \& {Messias}}]{Wielgus2022}
{Wielgus}, M., {Moscibrodzka}, M., {Vos}, J., {et~al.} 2022, \aap, 665, L6

\bibitem[{{Witzel} {et~al.}(2018){Witzel}, {Martinez}, {Hora}, {Willner},
  {Morris}, {Gammie}, {Becklin}, {Ashby}, {Baganoff}, {Carey}, {Do}, {Fazio},
  {Ghez}, {Glaccum}, {Haggard}, {Herrero-Illana}, {Ingalls}, {Narayan}, \&
  {Smith}}]{2018ApJ...863...15W}
{Witzel}, G., {Martinez}, G., {Hora}, J., {et~al.} 2018, \apj, 863, 15

\bibitem[{Witzel {et~al.}(2018)Witzel, Martinez, Hora, Willner, Morris, Gammie,
  Becklin, Ashby, Baganoff, Carey, Do, Fazio, Ghez, Glaccum, Haggard,
  Herrero-Illana, Ingalls, Narayan, \& Smith}]{Witzel2018}
Witzel, G., Martinez, G., Hora, J., {et~al.} 2018, \apj, 863, 15

\bibitem[{{Witzel} {et~al.}(2021){Witzel}, {Martinez}, {Willner}, {Becklin},
  {Boyce}, {Do}, {Eckart}, {Fazio}, {Ghez}, {Gurwell}, {Haggard},
  {Herrero-Illana}, {Hora}, {Li}, {Liu}, {Marchili}, {Morris}, {Smith},
  {Subroweit}, \& {Zensus}}]{Witzel2021}
{Witzel}, G., {Martinez}, G., {Willner}, S.~P., {et~al.} 2021, \apj, 917, 73

\bibitem[{{Yuan} {et~al.}(2009){Yuan}, {Lin}, {Wu}, \& {Ho}}]{Yuan2009}
{Yuan}, F., {Lin}, J., {Wu}, K., \& {Ho}, L.~C. 2009, \mnras, 395, 2183

\bibitem[{{Yuan} {et~al.}(2003){Yuan}, {Quataert}, \& {Narayan}}]{Yuan2003}
{Yuan}, F., {Quataert}, E., \& {Narayan}, R. 2003, \apj, 598, 301

\bibitem[{{Yusef-Zadeh} {et~al.}(2009){Yusef-Zadeh}, {Bushouse}, {Wardle},
  {Heinke}, {Roberts}, {Dowell}, {Brunthaler}, {Reid}, {Martin}, {Marrone},
  {Porquet}, {Grosso}, {Dodds-Eden}, {Bower}, {Wiesemeyer}, {Miyazaki}, {Pal},
  {Gillessen}, {Goldwurm}, {Trap}, \& {Maness}}]{Yusef-Zadeh2009}
{Yusef-Zadeh}, F., {Bushouse}, H., {Wardle}, M., {et~al.} 2009, \apj, 706, 348

\bibitem[{{Yusef-Zadeh} {et~al.}(2006){Yusef-Zadeh}, {Roberts}, {Wardle},
  {Heinke}, \& {Bower}}]{YusefZadeh2006}
{Yusef-Zadeh}, F., {Roberts}, D., {Wardle}, M., {Heinke}, C.~O., \& {Bower},
  G.~C. 2006, \apj, 650, 189

\bibitem[{{Zhu} {et~al.}(2019){Zhu}, {Li}, {Morris}, {Zhang}, \&
  {Liu}}]{2019ApJ...875...44Z}
{Zhu}, Z., {Li}, Z., {Morris}, M.~R., {Zhang}, S., \& {Liu}, S. 2019, \apj,
  875, 44

\bibitem[{Zubovas {et~al.}(2012)Zubovas, Nayakshin, \& Markoff}]{Zubovas_2012}
Zubovas, K., Nayakshin, S., \& Markoff, S. 2012, Monthly Notices of the Royal
  Astronomical Society, 421, 1315

\end{thebibliography}
\bibliographystyle{aa}

\appendix
\section{Ornstein-Uhlenbeck and AR(1) characteristic shape}\label{sec:ou_process}
The O-U process \citep{Uhlenbeck1930} a is a two-parameter, mean-reversing (stationary) process that generates correlated random motion and corresponds to a continuous damped random walk. It is the continuous equivalent to an AR process of first order, AR(1). \autoref{eqn:ou_defintion} gives the definition of the stochastic differential equation:
\begin{equation}
    dX_t = - \theta x_t dt + \sigma d W_t.\label{eqn:ou_defintion}
\end{equation}
which can be solved analytically \citep[e.g.,][]{Risken1989}:
\begin{equation}
    x_t = x_0 e^{-\theta t} + \mu(1-e^{-\theta t}) + \dfrac{\sigma}{\sqrt{2\theta}}W_{1-e^{-2 \theta t}}, \label{eqn:ou_process_solved}
\end{equation}
where $\mu$ stands for the process mean and $W_{1-e^{-2\theta t}}$ for the Wiener process. 

From this, the expectation value, $\mathbb{E}(x_t)$, can be derived, which depends linearly on the value of $x_0$ and otherwise decays exponentially with increasing time. Because of the linear dependence, $x_0$, the flares can be normalized, which always leads to the same expected decay (or increase), causing the observed exponential shape. This demonstrates that the observed exponential shape is indeed the expected value for an O-U process and thus relates the observed $\tau$ to the $\theta$ parameter of the process:

\begin{equation}
    \mathbb{E}(x_t) = x_0 e^{-\theta t} + \mu(1-e^{-\theta t}).\label{eqn:expectation_value}
\end{equation}

\begin{figure}
    \centering
    \includegraphics[width=0.485\textwidth]{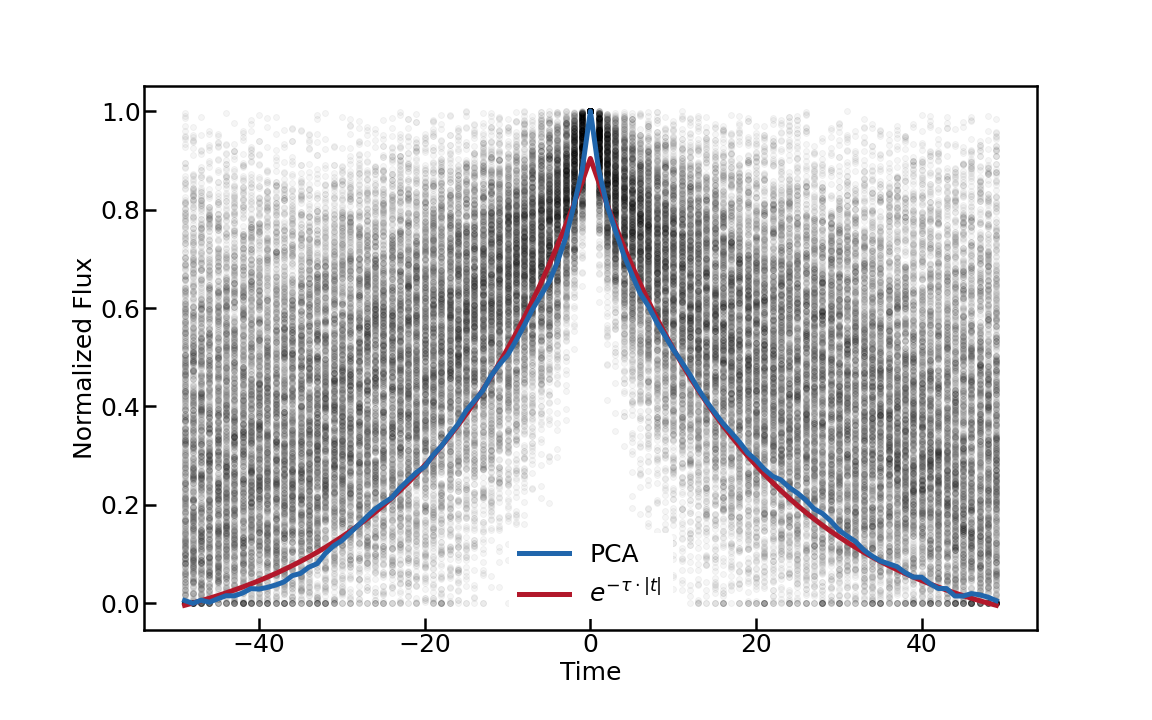}
    \caption{Exponential shape derived from selected, centering, and normalizing ``flares'' in a light curve generated by the O-U process (\autoref{eqn:ou_process_solved}).}
    \label{fig:ou_demo}
\end{figure}

In order to confirm this numerically, and in particular to test the ability to extract the relative quantities and to probe its biases, we applied our flare detection algorithm to simulated O-U-process light curves. \autoref{fig:ou_demo} shows the result, with the exponential shape expected from \autoref{eqn:expectation_value}. This also holds for exponentiated light curves (i.e., a log-OU process). We further confirm that the value derived by the PCA is insensitive to the $\sigma$ parameter of the O-U process and that it does not depend on the tuning parameters of the flare selection algorithm.



\section{PCA of the MA process}\label{sec:ma_pca}
\begin{figure*}
    \centering
    \includegraphics[width=0.985\textwidth]{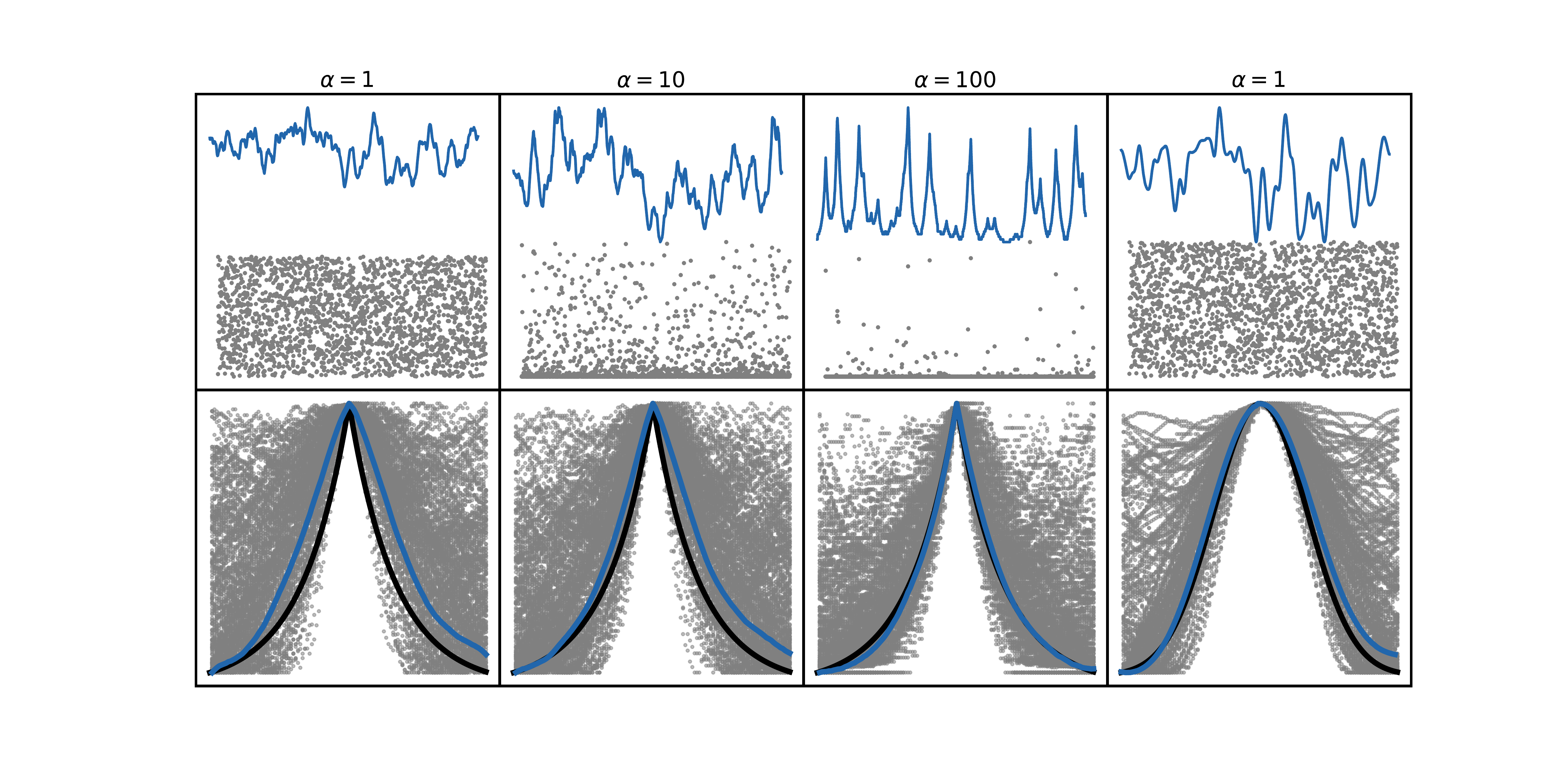}
    \caption{Light curves generated using the MA process and the same random seed, but with differing process parameters. Top panels: Light curve segment (blue line) and innovations $R(n) = \mathcal{U}^\alpha$ (gray points) for different values of $\alpha$. Bottom panel: Stacked, shifted, and normalized flares, defined as an isolated peak above the $80\%$ flux percentile (gray dots) in the light curve. The black line illustrates the impulse used in the light curve generation, and the blue shows the first component of a PCA of the data.}
    \label{fig:ma_process_pca}
\end{figure*}

In order to test the ability of PCA to extract intrinsic pulse profiles, we generated light curves from an MA process. Following the definitions in \cite{Scargle2020}, we generated discrete light curve values, $X(n),$ as
\begin{equation}
    X(n) = \sum_k c_k R(n-k) + D(n)\label{eqn:ma_process}
,\end{equation}
where $c_k$ are the impulse coefficients, and $R(\cdot)$ are $n$ random numbers with a ``white'' power spectrum. We generated $R(\cdot)$ using uniform random numbers $R \in (0,1]$, transformed by a power-law exponent, $\alpha$, and set the deterministic part of the light curve, $D(n)$, to zero. Depending on the value of $\alpha$, large variations occur (which we would interpret as flares; see \autoref{fig:ma_process_pca} and the discussion in \citealt{Scargle1981, Scargle2020}). Picking out high amplitude variations in the light curve and normalizing and stacking these segments reveals the impulse response used in the generation of the light curve (gray points in the lower panel of  \autoref{fig:ma_process_pca}). Remarkably, the PCA is able to pick out different impulse shapes; for example, one can differentiate between a Gaussian impulse and an exponential impulse. Even more remarkably, the PCA can approximate the intrinsic impulse in the case of $\alpha=1$, which \cite{Scargle2020} considered as the ``Gaussian limit in which the true flare shape cannot be determined by any algorithm, because the high degree of overlap hides information beyond second order statistics.''

In our tests we only explored a limited regime of the MA and impulse parameters that is comparable to our observations. In all cases, the PCA picked out the intrinsic impulse shape reasonably well, even for light curves that showed much less skewed flux distributions than what is observed for Sgr~A* \citep{GRAVITYCollaboration2020flux}. We argue that our procedure would pick out the intrinsic flare shape if the observed flux outburst could indeed be interpreted as such. In the next subsection, we explore the effect of (correlated) noise on the analysis.

\subsection{Noise biases in PCA}\label{sec:pca_noise_bias}
\begin{figure}

    \centering
    \includegraphics[width=0.5\textwidth]{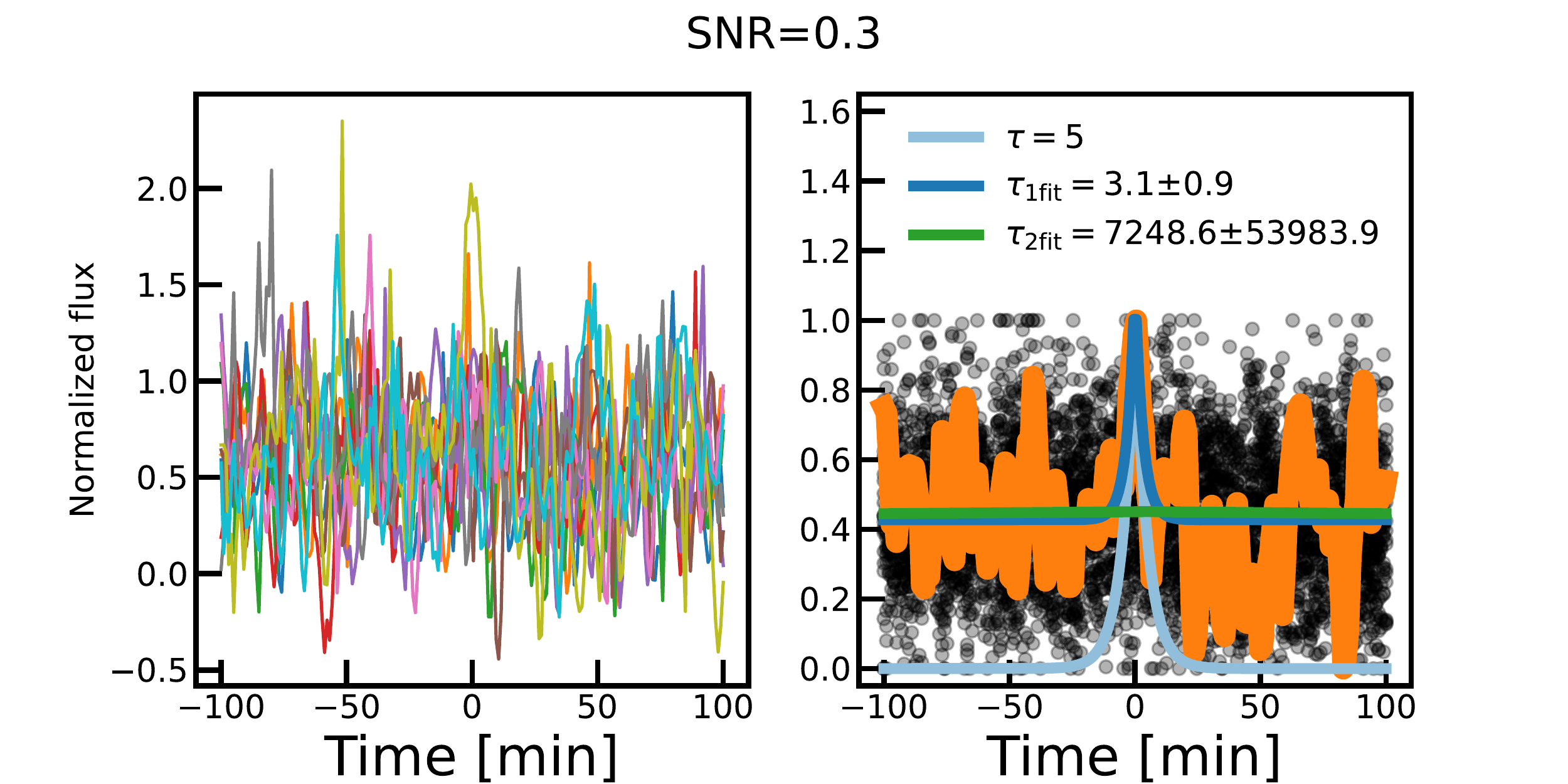}
    \caption{Response function detection using the PCA method in the presence of high correlated noise. We show the low-S/N scenario with $25$ flares, high gamma-distributed (Poissonian) noise, and strong correlated noise. Left: All $25$ flares as observed.\ Right: Normalized flares, overplotted with an intrinsic flare component (light blue) and the first PCA component (thick orange). Two fitted exponential functions are displayed, in dark blue an exponential function with offset and in dark green without. The median S/N of the flares (peak flux at $t=0$/ standard deviation) is $0.3$.}
    \label{fig:noise_bias_simulation}
\end{figure}

PCA decomposition works by constructing the orthogonal basis in which each component maximises the variances in the data. Thus, the derived components have no physical interpretation, and, when interpreted as such, care must be taken to avoid interpreting noise. In order to test the robustness of the PCA, we simulated flares with a known rise and decay time, a flux-dependent noise component (either Gaussian or Gamma-distributed), and a red-noise contribution. Exploring a wide range of parameters for the different noise components, we find that the input, $\tau$, is generally well recovered if one allows for an offset, even if the median S/N of the flares drops significantly below $1$. \autoref{fig:noise_bias_simulation} illustrates such a low-S/N scenario. Despite the low S/N, the intrinsic flare shape is recovered by the first PCA component and the derived rise and decay time, $\tau$, is recovered with $2\sigma$. Given the high S/N in the \textit{Spitzer} data, we are confident that we have recovered a physical meaningful component of the data. 

\section{Extreme cases of the hot-spot model}\label{sec:extreme_cases}
\begin{figure}
    \centering
    \includegraphics[width=0.485\textwidth]{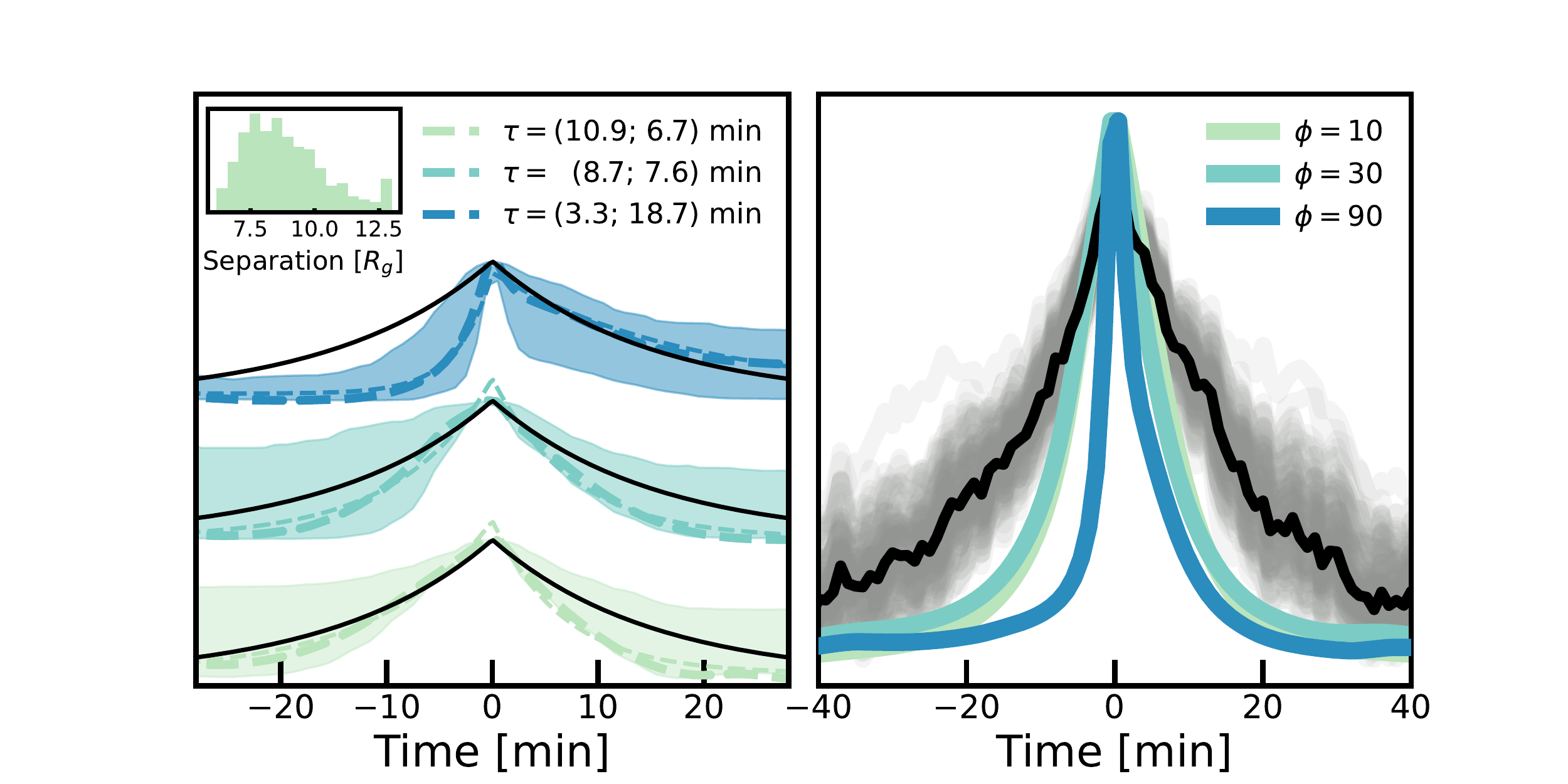}
    \caption{Extreme cases of the orbiting hot-spot model: purely exponential and purely constant intrinsic emission. Left: Extreme case of flares simulated with a constant intrinsic emission modulated by relativistic effects. We plot the $1\sigma$ envelope of the principal component fitted to $1000$ flare simulations (see the text for details). Three simulations are plotted, shifted by a constant amount to make them comparable. The bottom envelope shows the $i = 10\degree$ case, the middle envelope the $i = 50\degree$ case, and the top envelope the $i = 90\degree$ case. The figure inset shows the histogram of orbital separation used in the simulation. The legend gives the best fit rise and decay times of the exponential functions (thin dashed lines) fitted to the PCA component of the data (thick dashed). The black line shows the best fit exponential derived from the \textit{Spitzer} data. Right: Extreme case of flares simulated  with an intrinsic exponential flare shape with a rise and decay time of $\tau=\SI{15}{\minute}$, modulated by relativistic effects. Again, three scenarios for inclinations $i=10$, $i=50$, and $i=90$ are shown. The thick black line shows the PCA component derived from the \textit{Spitzer} data, and the thin gray lines show the bootstrapped surrogates, as in \autoref{fig:rise_decay_time}.}
    \label{fig:extrem_cases}
\end{figure}

In the first extreme case, we assumed a constant intrinsic flux level, which is modulated by the motion of the flare's relativistic orbit calculated from its separation, $R_n$. In this scenario, all flux variability is caused by the relativistic magnification of the otherwise constant light curve. We derived the first PCA component from the so-aggregated $1000$ light curves shown in the left panel of \autoref{fig:extrem_cases} for three different viewing angles. In all cases, the rising and falling flank can be approximated by an exponential (thin dashed line in \autoref{fig:extrem_cases}); however, except for with intermediate inclinations, the rise and fall times are unequal. Particularly in the case of a black hole viewed edge-on  ($\phi=90\degree$), the first component is asymmetric due to strong lensing magnification. Here, we chose a flare distribution according to a gamma distribution centered on $8R_g$, which leads to generally shorter rise and decay times. For an approximately symmetric exponential flare shape with $\tau=\SI{15}{\minute,}$ the flares would need a larger radial separation, which is not accessible with our relativistic kernel.

In the second extreme case we assumed an intrinsic flare shape corresponding to an exponential of $\tau=\SI{15}{\minute}$, which is modulated by the relativistic effects of the flare's orbit around the black hole. Importantly, the relativistic modulation in all cases leads to a narrower profile than the input exponential. Further, very high inclinations lead to an asymmetric profile inconsistent with the observed value.

\section{Flare fit overview}\label{sec:flarefit_overview}
\begin{figure*}[b]
    \centering
    \includegraphics[width=0.985\textwidth]{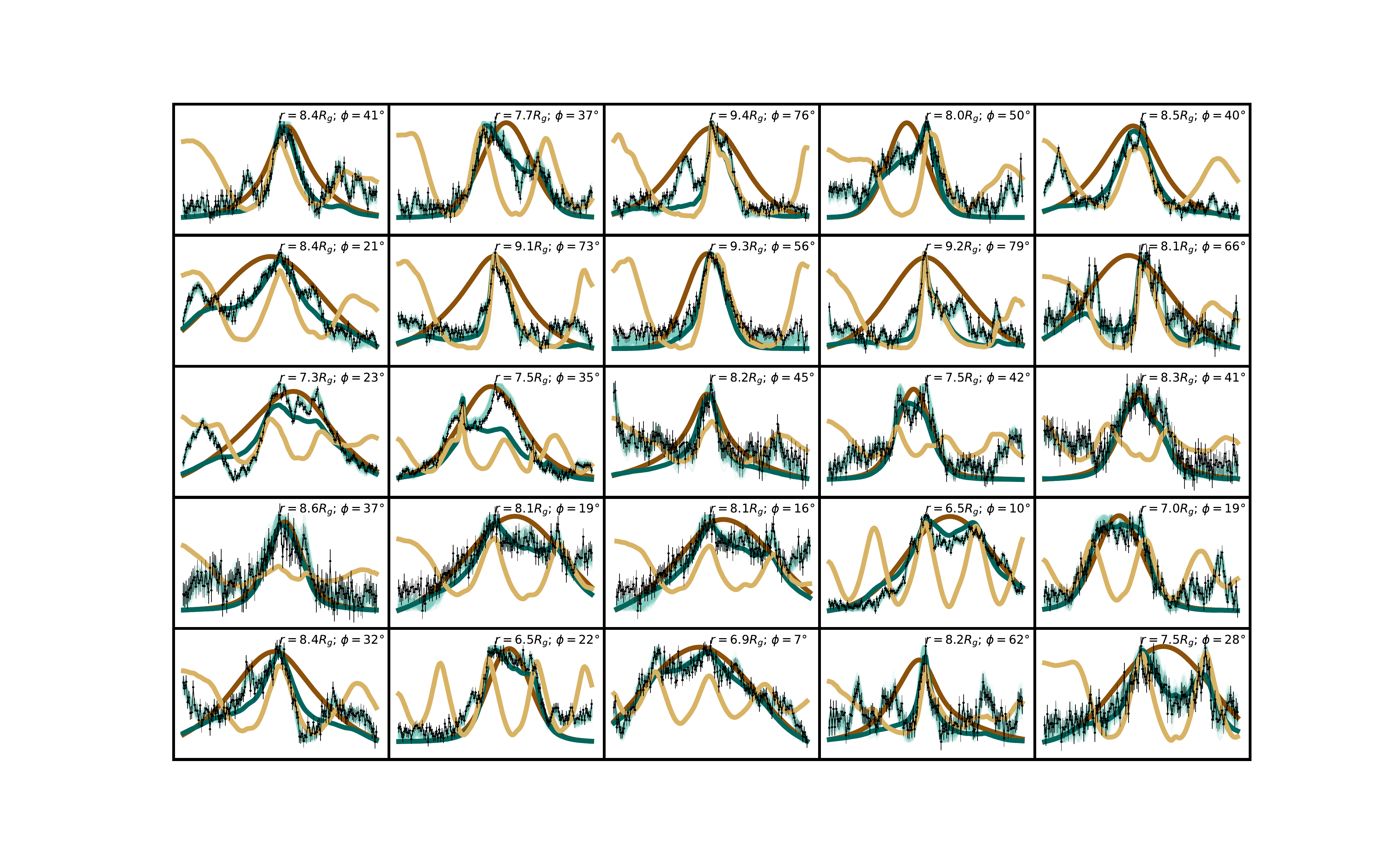}
    \caption{Flares and posterior samples of the orbiting hot-spot model. This figure is similar to \autoref{fig:model_fit_example}, but shows the fits to all 24 flares. The models shown are not the best fit but rather the average of $100$ models drawn from the posterior. This illustrates features of the multimodal posteriors. The light yellow line shows the averaged relativistic kernel, the dark brown shows the averaged Gaussian flare kernel, the dark green shows the composite model without the Gaussian process component, and the light green shows the full model. The annotated values of $\phi$ and $r0$ are median posterior values.}
    \label{fig:all_flare_fit}
\end{figure*}

\autoref{fig:all_flare_fit} shows the fit results for all $25$ flares. 

\end{document}